\documentclass{aa}  
\usepackage{booktabs}
\usepackage{longtable}
\usepackage{mathabx}
\usepackage{graphicx}
\def\kms{\hbox{km$\;$s$^{-1}$}}

\usepackage{txfonts}
\usepackage{natbib,twoopt}

\begin{document}


   \title{Semi-empirical model atmospheres for the chromosphere of the sunspot penumbra  and umbral flashes}


   \author{Souvik Bose
          \inst{1}$^,$\inst{2},
          Vasco M.J. Henriques\inst{1}$^,$\inst{2},
          Luc Rouppe van der Voort\inst{1}$^,$\inst{2},
          \and 
          Tiago M.D. Pereira\inst{1}$^,$\inst{2}
          }

   \institute{Institute of Theoretical Astrophysics, University of Oslo, 
   P.O. Box 1029 Blindern, NO-0315 Oslo, Norway
         \and
             Rosseland Centre for Solar Physics, University of Oslo, P.O. Box 1029 Blindern, NO-0315 Oslo, Norway\\
             \email{souvik.bose@astro.uio.no}
             }

   \date{Received15.02.2019; accepted 20.05.2019}

 
  \abstract
   {The solar chromosphere and the lower transition region is believed to play a crucial role in the heating of the solar corona. Models that describe the chromosphere (and the lower transition region), accounting for its highly dynamic and structured character are, so far, found to be lacking. 
   This is partly due to the breakdown of complete frequency redistribution (CRD) in the chromospheric layers and also because of the difficulty in obtaining complete sets of observations that adequately constrain the solar atmosphere at all relevant heights.}
   {We aim to obtain semi-empirical model atmospheres that reproduce the features of the \ion{Mg}{ii}~h\&k line profiles that sample the middle chromosphere with focus on a sunspot.}
   {We use spectropolarimetric observations of the \ion{Ca}{ii}~8542~\AA\ spectra obtained with the Swedish 1-m Solar Telescope and use NICOLE inversions to obtain semi-empirical model atmospheres for different features in and around a sunspot. These are used to synthesize \ion{Mg}{ii}~h\&k spectra using RH1.5D code, which we compare with observations taken with the Interface Region Imaging Spectrograph (IRIS).}
   {Comparison of the synthetic profiles with IRIS observations reveals that there are several areas, especially in the penumbra of the sunspot, where most of the observed \ion{Mg}{ii}~h\&k profiles are very well reproduced. In addition, we find that supersonic hot downflows, present in our collection of models in the umbra, lead to synthetic profiles that agree well with the IRIS \ion{Mg}{ii}~h\&k profiles, with the exception of the line core.}
   {We put forward and make available four semi-empirical model atmospheres. Two for the penumbra, reflecting the range of temperatures obtained for the chromosphere, one for umbral flashes, and a model representative of the quiet surroundings of a sunspot. These are available in electronic as well as in table formats.}

  \authorrunning{Bose et. al}
  \titlerunning{Semi-empirical model atmospheres}
   \maketitle
%

\section{Introduction}
The solar chromosphere is of much interest to the solar physics community because of its highly dynamic nature and also due to the crucial role it is believed to play in the heating of the solar corona. It is sandwiched between the photosphere and the corona and
its density decreases radially outwards from the surface of the Sun, whereas the temperature decreases from the photosphere-chromosphere boundary, reaches a minimum and then undergoes a drastic increase into the transition region \citep[see][]{Val1981,Avrett2003}.

The study and analysis of the chromosphere is challenging as well as rewarding. Challenging because of its complex spatial and temporal structures that are not completely resolved by the present state-of-the-art instrumentation; and rewarding because the energy that leads to coronal heating, the magnetic flux that passes through it to form the coronal magnetic field, is thought be taking place in the chromosphere \citep[e.g.][]{Jaime2016}. Further, the spectral lines that appear in this region are strictly formed under optically thick non-Local Thermodynamic Equilibrium (non-LTE), which makes the interpretation and analysis rather difficult. Nevertheless, there have been various successful attempts to interpret chromospheric observations using non-LTE inversions. Some of the available non-LTE inversion codes are \verb|HAZEL| \citep{Ramos2008}, \verb|HELIX| \citep{Lagg2009}, \verb|NICOLE| \citep{Hector2015}, \verb|STiC| \citep{2019A&A...623A..74D} and \verb|SNAPI| \citep{Milic2018}. The aim behind all of these codes is to obtain a best-fit of the observed chromospheric profiles by iteratively modifying the atmospheric parameters. 

Seated above the chromosphere, is a thin region where the temperature rises drastically from 0.02 MK to about 0.8 MK and the density falls off rapidly. This layer is called the solar transition region (TR). It transitions between collisionally dominated and partially ionized plasma, and collision-less, fully ionized plasma. One-dimensional solar atmospheric models such as that of \citet{Val1981} predict that this sharp temperature rise takes place within 100 km in the solar atmosphere. However, recent spectroscopic and imaging observations from the \textit{Interface Region Imaging Spectrograph} \citep[IRIS:][]{Bart2014}, reveal that the TR is very structured and inhomogeneous in nature. Most of the spectral lines that form in this region fall in the range between near ultraviolet ($\approx 3000$ \AA\,) and far ultraviolet ($\approx$ 400 \AA\,) where the continuum opacity is very high.

The IRIS \ion{Mg}{ii}~h\&k spectral lines in particular, provides us with one of the richest diagnostics of the upper chromosphere and lower TR \citep{Leenaarts2013b,Tiago2015,Jaime2016}. Results from \citet{Leenaarts2013a,Leenaarts2013b,2017A&A...597A..46S} reveal that these lines are strictly non-LTE and the k$_\mathrm{2}$ \& h$_\mathrm{2}$ emission peaks are affected by partial frequency redistribution (PRD). These peaks are representative of the conditions that exist in the middle and in the upper-middle chromosphere. 
The k$_{3}$ \& h$_{3}$ line cores (formed higher up) are also partially affected by PRD \citep{2017A&A...597A..46S}. Further, 3D radiative transfer effects needs to be accounted for \citep{Leenaarts2013a,2017A&A...597A..46S} to model these cores. However, this effect is smaller for higher $\mu$ of the observations (with $\mu$ being the cosine of the heliocentric angle) and the difference between 3D and 1D radiative transfer is often negligible \citep[as is the case with FAL-C modified with different velocity fields,][]{2017A&A...597A..46S}.  

\citet{Leenaarts2013b} note that the \ion{Ca}{ii}~K and \ion{Mg}{ii}~k cores have very similar formation heights, and are formed significantly higher than \ion{Ca}{ii}~8542 \AA\, and H$\alpha$. 
Both \ion{Ca}{ii}~H\&K and \ion{Mg}{ii}~h\&k lines have similar oscillator strengths. However, because Mg is about 18 times more abundant than Ca \citep{Asplund2009}, the h\&k lines are stronger and therefore form slightly higher than the H\&K lines. 

While there exist several studies on simultaneous observations of the chromosphere and the TR, semi-empirical atmospheric models of a sunspot, that effectively characterize the chromosphere and the lower TR are found to be rather uncommon, especially for the penumbra. \citet{Socas-Navarro2007models} presented the first chromospheric semi-empirical models for a sunspot, where they used non-LTE inversions of four \ion{Ca}{ii} and \ion{Fe}{i} spectra. They include the only chromospheric models for the penumbra till date. However, the uncertainties in the chromosphere are quite high, perhaps due to the difficulty in properly constraining chromospheric penumbral models. 

In this paper, we aim to obtain semi-empirical model atmospheres that reproduce the line profiles formed in the chromosphere of a sunspot, especially the penumbra and the umbral flashes. An empirical understanding of these atmospheres will help us to set better observational estimates of this highly dynamic region and will further allow us to obtain the height dependencies of various physical parameters like temperature, hydrogen populations, microturbulence, line-of-sight (LOS) velocities, and electron density from the solar surface. Spectral inversions of these layers in sunspots have so far proven difficult because semi-empirical models, used as starting atmospheres, did not account for the effects such as PRD. 

In this regard, we focus on the chromospheric model atmospheres obtained with \verb|NICOLE| inversions of the \ion{Ca}{ii}~8542 \AA\, line observed with the Swedish 1-m Solar Telescope \citep[SST:][]{Scharmer2003SST}, and then synthesize upper chromospheric line profiles, such as \ion{Mg}{ii}~h\&k, using RH1.5D \citep{Tiago2015RH} with the chromospheric atmospheres as a starting point. Further, the synthetic profiles are compared with the observed IRIS spectra to further empirically test the validity of selected models. This approach serves as a consistency check on these semi-empirical atmospheres, and adequately constrains the solar atmosphere at all heights for the highly dynamic small-scale features of a sunspot.

This paper is divided into the following Sects: Sect.~\ref{sec:obs} describes the data we used in this paper. In Sect.~\ref{sec:methods} we discuss the methodology, followed by Sect.~\ref{sec:results}, where we describe the results. Finally, we finish off with discussions and conclusions in Sect.~\ref{sec:conclusion}. 

\section{Observations}
\label{sec:obs}

We used observations of the active region NOAA 12533 in a coordinated campaign between the SST and IRIS on 29 April 2016 starting from 09:42 UT, with solar $(x, y)$ coordinates of $(623\arcsec, 19\arcsec)$ and $\mu$=0.745. We used the CRisp Imaging SpectroPolarimeter \citep[CRISP:][]{Scharmer2008crisp} to record the data with SST in imaging spectropolarimetric mode. CRISP is a dual etalon Fabry-Pérot interferometer with low resolution etalons in a telecentric beam configuration. The Multi-Object Multi-Frame Blind Deconvolution \citep[MOMFBD:][]{vannoort2005MOMFBD} technique was used in the reduction of the raw data. The CRISP data reduction pipeline was used for further reduction as in \citet{Jaime2015CRISPRED} with the small-scale seeing consistency method for spectral profiles as described in \citet{Vasco2012}. CRISP sampled the \ion{Ca}{ii}~8542~\AA\  line at 21 wavelength positions between $\pm$ 1.75~\AA\ with respect to the line core, in spectropolarimetric mode. IRIS observed in a medium sparse 8-step raster mode with 4~s exposure time and 1\arcsec raster steps (OBS-ID 3620106129). The \ion{Ca}{ii}~8542~\AA\  line-profiles were normalized to disk-center quiet Sun continuum levels as measured in the atlas of \cite{1987brault_FTSatlas,1999SoPh..184..421N} following the procedure of  \cite{Jaime2013_Flashes}. The calibration factor was found by fitting a $\mu$-angle compensated average quiet-Sun profile, from a region away from the sunspot, to the atlas convolved with CRISP's transmission profile. Compensation for the $\mu$ angle was done using the tables from \cite{2011A&A...528A.113D}.

\begin{figure*}
   \centering
   \includegraphics[width=10.5cm,angle=90]{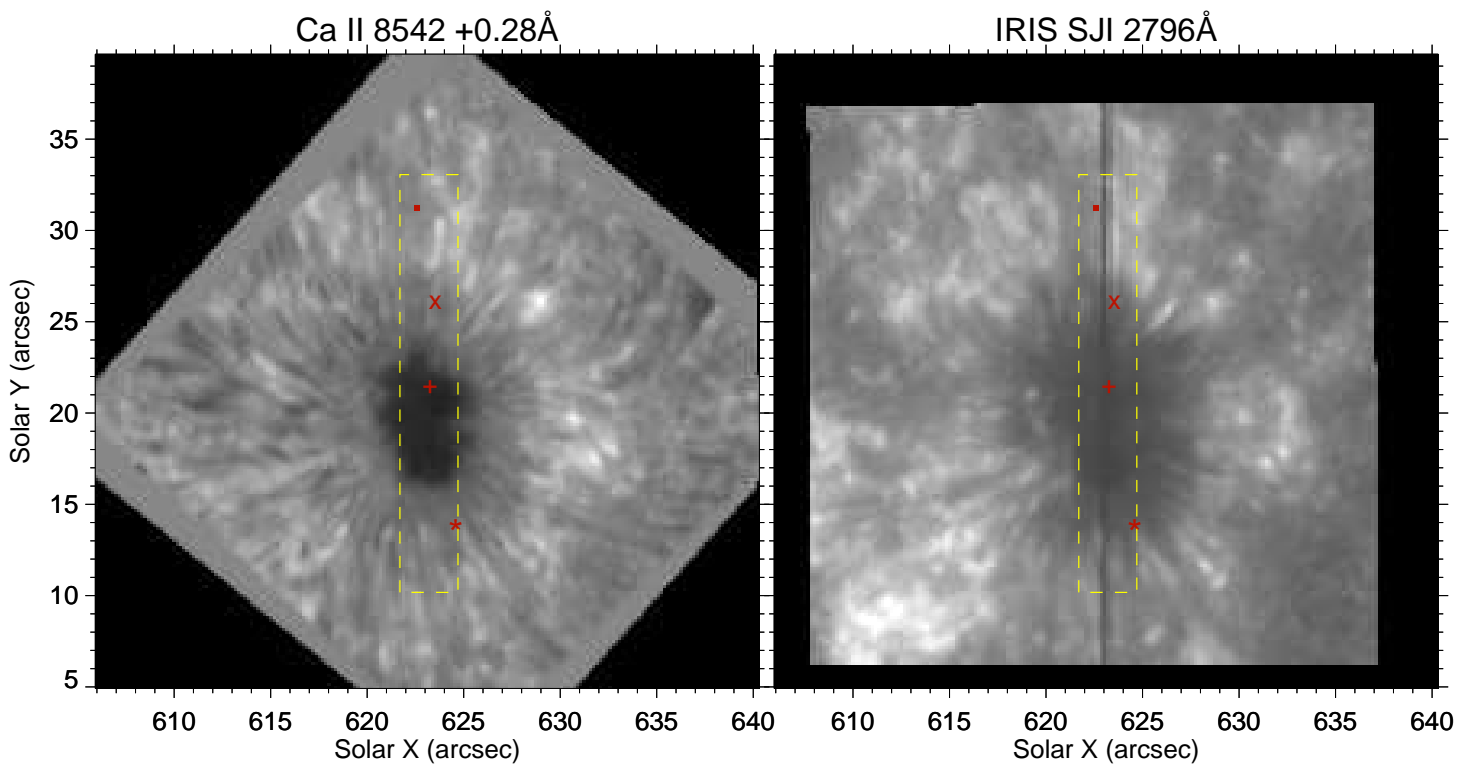}
   \caption{Full FOV of the scans from the SST and IRIS on the 29th April 2016, at 09:55 UT. Tick-marks in solar coordinates. Left panel: \ion{Ca}{ii}~8542 \AA\, red wing image obtained from CRISP. Right panel: Corresponding IRIS SJI at 2796 \AA\,. The yellow dashed region shows the inverted FOV for reference. The symbols: plus sign, cross, asterisk, and square, respectively correspond to selected locations featuring an umbral flash, the cool penumbra model, the hot penumbra model and a quiet surrounding pixel, respectively. The atmospheric models corresponding to these regions are provided in this paper.}
              \label{fig:context}
    \end{figure*}

The CRISP data, with a spatial sampling of 0\farcs058 per pixel and diffraction limit $\lambda/D$=0\farcs18 at 8542 \AA, was rotated, aligned and re-sampled to the IRIS Slit-Jaw (SJI) \ion{Mg}{ii} 2796~\AA, data. The co-temporal and co-spatial field-of-view (FOV) of the data was $34.4\times34.76$ arsec$^{2}$ with a final pixel scale of about 0\farcs167~pixel$^{-1}$. For the current work, we considered a single scan of the data obtained at 09:55 UT along with the IRIS observations as shown in Fig.~\ref{fig:context}.

\section{Methods and Analysis}

\label{sec:methods}

\subsection{NICOLE Inversions}

We performed spectropolarimetric inversions of the \ion{Ca}{ii}~8542~\AA\ data over the region enclosed by the yellow dashed line shown in Fig.~\ref{fig:context} using the \verb|NICOLE| code \citep{Hector2015}. The inverted FOV covers an area of $3.2\times23$ arcsec$^{2}$ (i.e. $19\times138$ pixel$^{2}$).  \verb|NICOLE| is a multi-purpose inversion code that is parallelized for both synthesis and inversions of Stokes profiles under non-LTE conditions. It requires an initial guess atmosphere for calculating the emergent profiles. In inversion mode, the code iteratively modifies the different physical parameters such as temperature, velocity, and microturbulence of the initial guess atmosphere using a Marquardt nonlinear least-squares minimization algorithm incorporating response functions \citep[as in][]{1992ApJ...398..375R} to reach a match between the synthetic and the observed profiles. The chosen initial atmosphere was the quiet-sun model commonly referred to as \verb|FAL-C|, from \citet{FALC1993}, for all pixels in the FOV.

The \ion{Ca}{ii} ion was modelled as a five bound-level plus continuum \ion{Ca}{ii} atom, as used by  \cite{Leenaarts2009}. \verb|NICOLE| operates under complete angle and frequency redistribution  \citep[CRD,][]{ScharmerandCarlsson85}, which is a very good approximation for the \ion{Ca}{ii} Infrared lines \citep{1989A&A...213..360U}. The starting LTE populations were determined using the \verb|MULTI| approach, as implemented by \citet{Carlsson1986MULTI}, where the sum of statistical weights in the Saha equation is done over the finite number of levels of the model atom. A Gaussian quadrature with 3 angles was selected for the rays.

 The cubic DELO-Bezier formal solver \citep{Jaime2013_Bezier} was chosen for the radiative transfer equation. Following \cite{Leenaart2014}, we included the \ion{Ca}{ii} isotopes to obtain good fits to the observations in the red wing of the profiles. 

Nodes were placed regularly at equidistant heights from the topmost optical depth of $\log\tau_{5000} = -8$ to approximately $\log\tau = 1$. We used the native \verb|NICOLE| equation-of-state. The electron pressure at the topmost height, for all pixels remained unchanged by the inversion cycles, and was kept at a value of 10$^{-1}$~dyn~cm$^{-2}$ as an upper boundary condition. Hydrostatic equilibrium is then used to stratify the atmosphere. Five iterative inversion cycles, where the results from each cycle were used as starting guesses to the next cycle, were used to improve convergence and avoid local minima when minimizing the ${\chi}^2$. The number of nodes per cycle are shown in Table~\ref{T1}. We performed multiple tests with lower nodes in temperature and LOS velocity, which led to significantly worse fits, especially for umbral flashes. Tests with one node in LOS magnetic field  led to the same models as averaging the three node inversions in the way described in Section 4.3 but, paradoxically, poorer matches for Stokes Q and U for otherwise well fitted penumbra profiles. Perturbations in microturbulence are additive, applied across the whole atmosphere, to the  stratification of the initial guess atmosphere (FAL-C). For cycle 3 to 5 the weights of Stokes V for the merit function were half of those for Stokes I. For Stokes Q and U the weights were four times lower than those of Stokes I. For cycle 2 these relations were double as biased in favor of Stokes I. This was to account for the different levels of noise and guarantee that the correct valley in the minimization of ${\chi}^2$ was selected early on. The comparison of the observed and the inverted Stokes parameters, corresponding to the selected locations of a hot penumbra, a  cool penumbra, and an umbral flash (as indicated in Fig.~\ref{fig:context}), is shown in the first four rows of Fig.~\ref{figure:stokes_and_models}. The last two rows of the same figure show the stratification of the respective model atmospheric parameters such as the temperature, LOS velocity and magnetic field, and micro-turbulent velocity, as derived from the \verb|NICOLE| inversions. 
\begin{table}
\caption{Number of nodes for different parameters}            
\label{T1}   
\centering                          
\begin{tabular}{c c c}       
\hline\hline                 
Nodes cycle 1 & Nodes cycle 2 to 5 & Parameters \\
\hline                        
   5 & 8 &  Temperature \\
   3 & 5 & Velocity \\ 
   1 & 1 & Microturbulence \textsuperscript{a} \\  
   0 & 1 & $B_{z}$ \\ 
   0 & 1 & $B_{x}$ \\ 
   0 & 1 & $B_{y}$ \\
   0 & 0 & Macroturbulence\\
\hline     
\end{tabular}

\footnotesize{ \textsuperscript{a)} Perturbations in microturbulence are additive, applied to the stratification of the initial atmosphere}
\end{table}

\subsection{Radiative Transfer using RH1.5D}

The RH1.5D\footnote{The code is publicly available at \url{https://github.com/ITA-Solar/rh}} \citep{Tiago2015RH} is a massively parallel radiative transfer code based on RH \citep{Uitenbroek2001}. It is capable of solving multi-level, multi-atom, non-LTE calculations by considering PRD and can also include Zeeman splitting effects allowing full Stokes synthesizing capabilities. We use the fast angle hybrid approximation for the PRD calculations as described in \citet{Leenaarts_FAST_angle}. This code can compute the spectra from either a 3D/2D/1D model atmosphere on a column-by-column basis (hence 1.5D) and it is designed to run over multiple nodes in a supercomputing cluster. 

The column-by-column approach, though faster, has one major limitation. It can only calculate the emergent profile vertically along the column. This means that the effect of inclined rays, crossing different atmospheres, is neglected, an effect that is possible to account for in a true 2D or 3D calculation. Further, the non-inclusion of the inclined rays might lead to a different mean radiation field, thereby affecting the non-LTE source function and ultimately the emergent intensity profile that may result in unexpected cooler cores of some very strong spectral lines \citep{Leenaarts2013a,Leenaarts2013b}. This can also lead to a decreased RMS contrast, compared to the column-by-column approach, at the line core of some spectra such as the \ion{Ca}{ii}~8542 \AA\ as shown by \cite{Leenarts_Carlsson_ca8542}. However recently, \citet{Jaime_2017_review} (and references therein) suggested that 1.5D geometry suffices for modelling the Ca infrared triplet lines. Nevertheless, for most non-LTE computations the 1.5D approach suffices, and does a reasonably good job. Besides, the ability of treating the spectra in PRD outweighs the drawbacks  significantly.

\subsection{Synthesizing \ion{Mg}{ii}~h\&k spectra.}

\label{para:RH synthetic spectra}
One of the major goals of this paper is to provide semi-empirical model atmospheres of a sunspot that are well constrained from the photosphere to the chromosphere, to obtain a pool of inverted models from NICOLE that can satisfactorily reproduce the \ion{Mg}{ii}~h\&k profiles. In this regard, we performed synthesis of the \ion{Mg}{ii}~h\&k spectra under PRD (using RH1.5D) with the set of model atmospheres obtained from the \verb|NICOLE| inversions of the SST data. We compared these synthetic profiles with the actual co-temporal and co-spatial observations from IRIS, that effectively serve as a check on the quality and the reliability of these semi-empirical models. 

We used a 10 level plus continuum model of the Mg atom for the radiative transfer computations. In observations, the \ion{Mg}{ii}~h\&k lines have blends from various other atomic species besides \ion{Mg}{ii}, and the most appropriate way would be to synthesize the spectra from RH by including as many atoms as possible in non-LTE. However, this adds tremendously to the time consumption of the synthesis. To circumvent the problem, we treated those atoms in LTE and used a line list that contained a list of bound-bound transitions and their parameters. RH supports line-lists in Kurucz format (as described in \url{http://kurucz.harvard.edu/linelists.html}). Once the spectra are synthesized, they were spectrally smeared by convolving with the IRIS instrumental profile to compare with the observations \citep[see][]{2013ApJ...778..143P}.

\section{Results and Discussion}
\label{sec:results}

\begin{figure*}
   \centering
   \includegraphics[width=\textwidth,height=23cm]{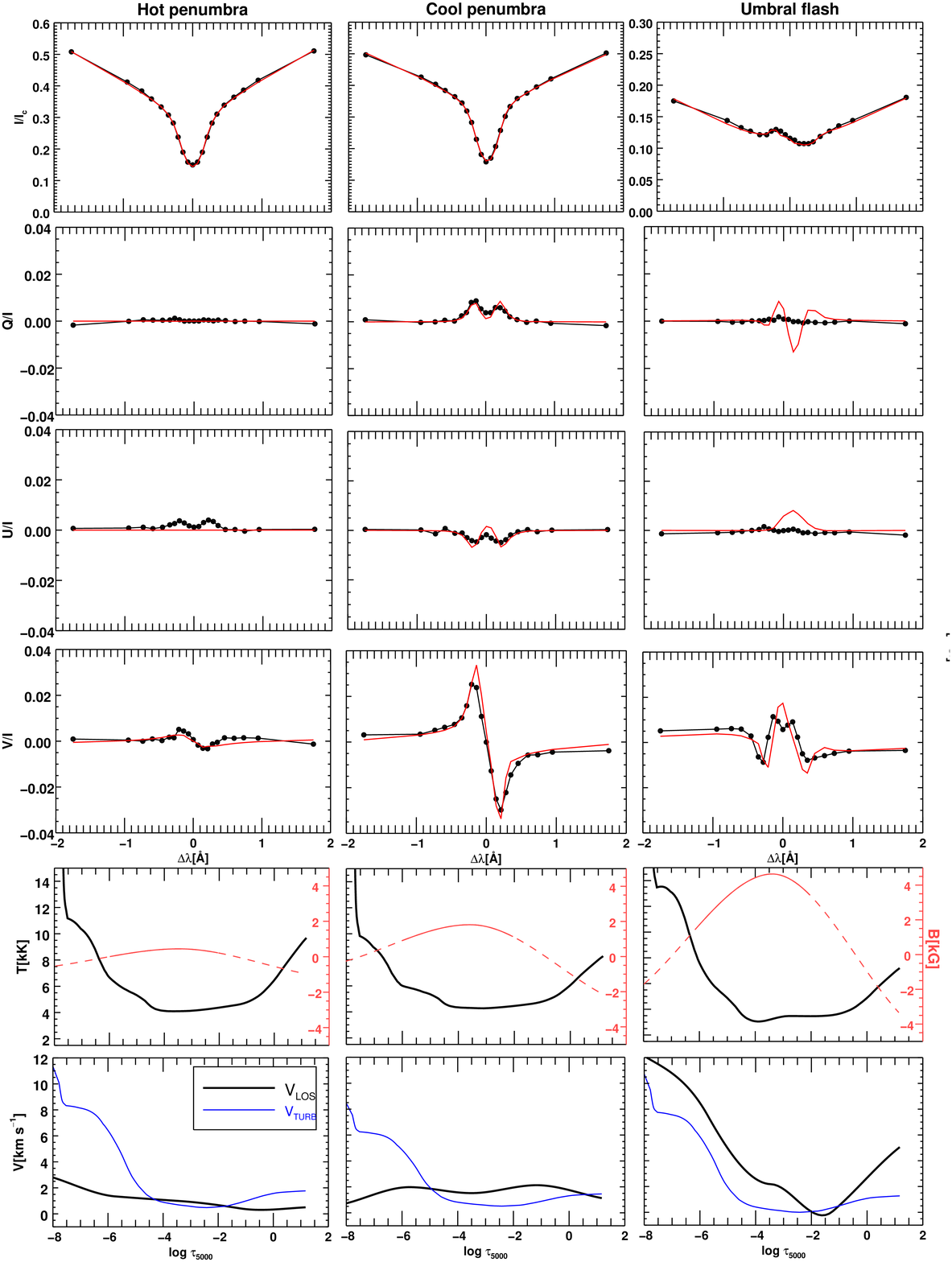}
   \caption{Columns showing the observed (black dots) and synthetic (red) full Stokes spectra in units of normalized HSRA \citep{HSRA1971} continuum intensity at disk center at a wavelength in the middle of the spectral range, and their atmospheric parameters such as temperature, LOS magnetic field,  velocity, and microturbulent velocity for three different models: hot penumbra (asterisk), cool penumbra ('cross') and umbral flash ('plus sign'), respectively. The continuous line overplotted on the dashed line for the LOS magnetic field, shows the variation of the magnetic field in the region $\log\tau =[-6,-2]$ where the Ca~8542 spectra is most sensitive. }
              \label{figure:stokes_and_models}%
    \end{figure*}

\subsection{Model atmospheres }
The three columns in Fig.~\ref{figure:stokes_and_models} show the observed and the inverted Stokes parameters along with the model atmospheric parameters for a hot penumbra, a cool penumbra, and an umbral flash respectively, obtained from the \verb|NICOLE| inversions. The different parameters shown are temperature, LOS velocity, LOS magnetic field and the microturbulent velocity. The cool and hot penumbra selection is representative of the range of temperature stratification obtained between $\log\tau = -5$ and $-$7, where the atmospheres show a divergence when compared to other layers. The multiple nodes in temperature, LOS velocity, and LOS magnetic field, as indicated in Table~\ref{T1}, are used to capture the variations in the vertical stratification of the solar atmosphere, including those of the upper chromospheric layers, as made possible by the different formation heights of the different wavelengths sampled. Atmospheric models such as these are used as inputs for the RH1.5D radiative transfer code for synthesizing the \ion{Mg}{ii}~h\&k spectra. The occasional extrapolation performed by \verb|NICOLE|, or the occasional large change in a node where the response in the \ion{Ca}{ii}~8542 line profile is relatively limited by the observations, is considered, for the purpose of this approach, as an advantage. Extrapolations as such led to a pool of models that work well in reproducing the observed properties of the \ion{Mg}{ii}~h\&k line profiles.

\subsection{Comparison between the synthetic and IRIS spectral profiles}
\label{subsection:comparision}
In Fig.~\ref{figure:rh_vs_iris_spectra} we show the comparison between the \ion{Mg}{ii}~h\&k profiles synthesized with RH1.5D (black) and the ones observed with the IRIS spectrograph (red). We show spectra of nine pixels (including the four marked in Fig.~\ref{fig:context}) spread over the FOV, selected based on the $\chi^{2}$ values between the observed and the synthetic profiles for the penumbra, and manually for the umbral flash and the quiet surrounding regions. We have labeled them from A through I corresponding to different solar features such as the penumbra, umbra and the relatively quiet areas away from the spot (termed as quiet surroundings). The synthetic profiles were spectrally smeared by convolving with a Gaussian profile, as described in Sect.~\ref{para:RH synthetic spectra}, to match the observations. Both the observed and the synthetic profiles were normalized to a reference intensity in the wavelength region between 2800~\AA\ and 2802~\AA\ averaged over a quiet area covering $0\farcs834\times1\farcs67$, located away from the sunspot. Further, we show the fits to the \ion{Ca}{ii}~8542 Stokes I profiles in Fig.~\ref{figure:nicole_inv_observed}, corresponding to the pixels shown in Fig.~\ref{figure:rh_vs_iris_spectra}, to indicate the excellent quality of the fits obtained from inversions.
\subsubsection{Penumbral profiles}
\begin{figure*}
   \centering
   \includegraphics[width =\textwidth]{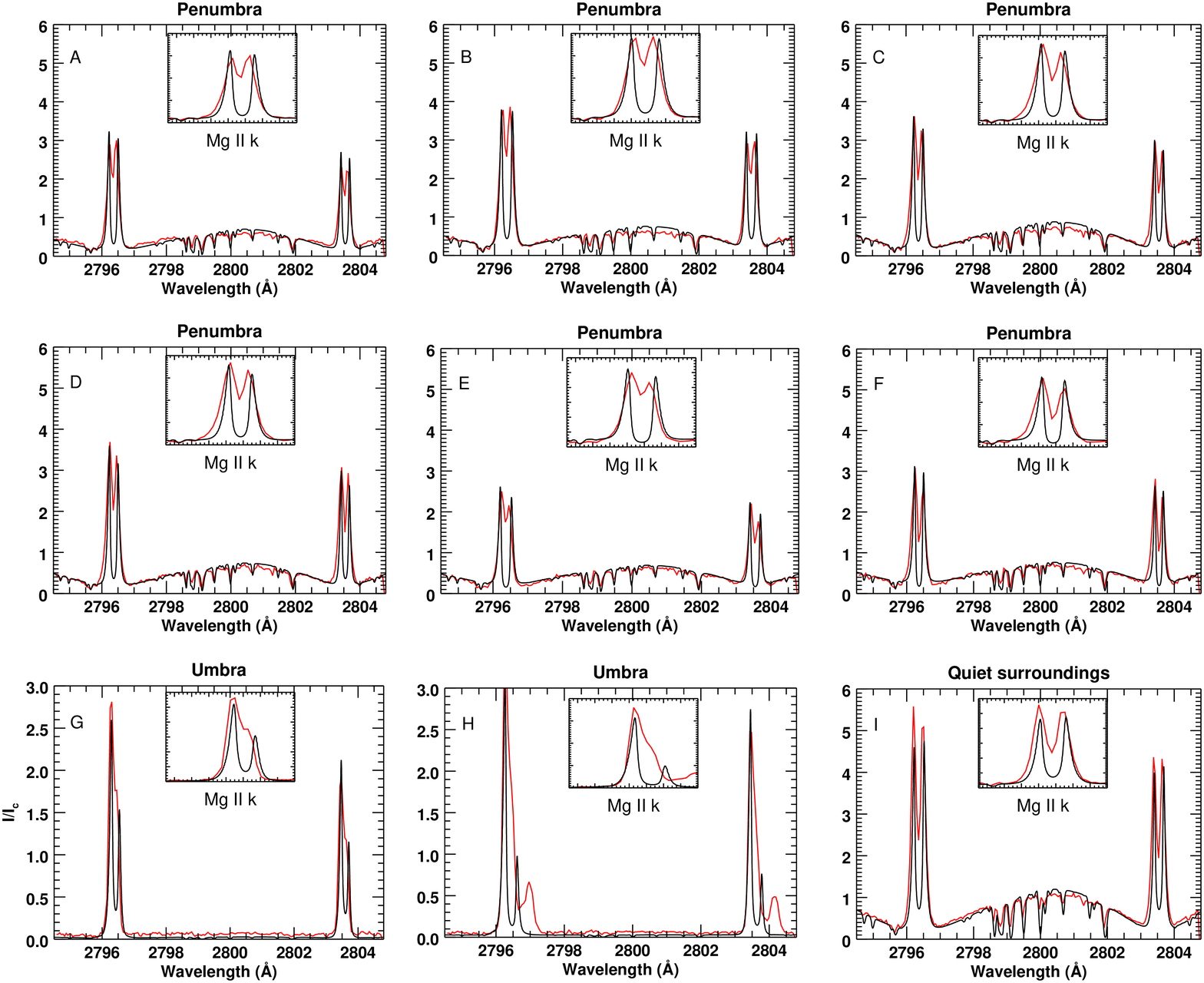}
   \caption{Comparison of the synthesized \ion{Mg}{ii}~h\&k spectra (black) with co-spatial and co-temporal IRIS observations (red) for various features on the FOV like penumbra, umbra and quiet surroundings. Profiles (A), (E), (G), and (I) correspond to the hot penumbra, cool penumbra, umbral flash and quiet surrounding respectively, with atmospheric models described in the previous section. The insets in each of the subplots zooms in the profiles for the \ion{Mg}~{\sc ii}~k region. Both the IRIS and the RH profiles were normalized to a reference intensity (I$_{c}$) as described in Section~\ref{subsection:comparision}.}
              \label{figure:rh_vs_iris_spectra}%
    \end{figure*}

\begin{figure*}
   \centering
   \includegraphics[width =\textwidth]{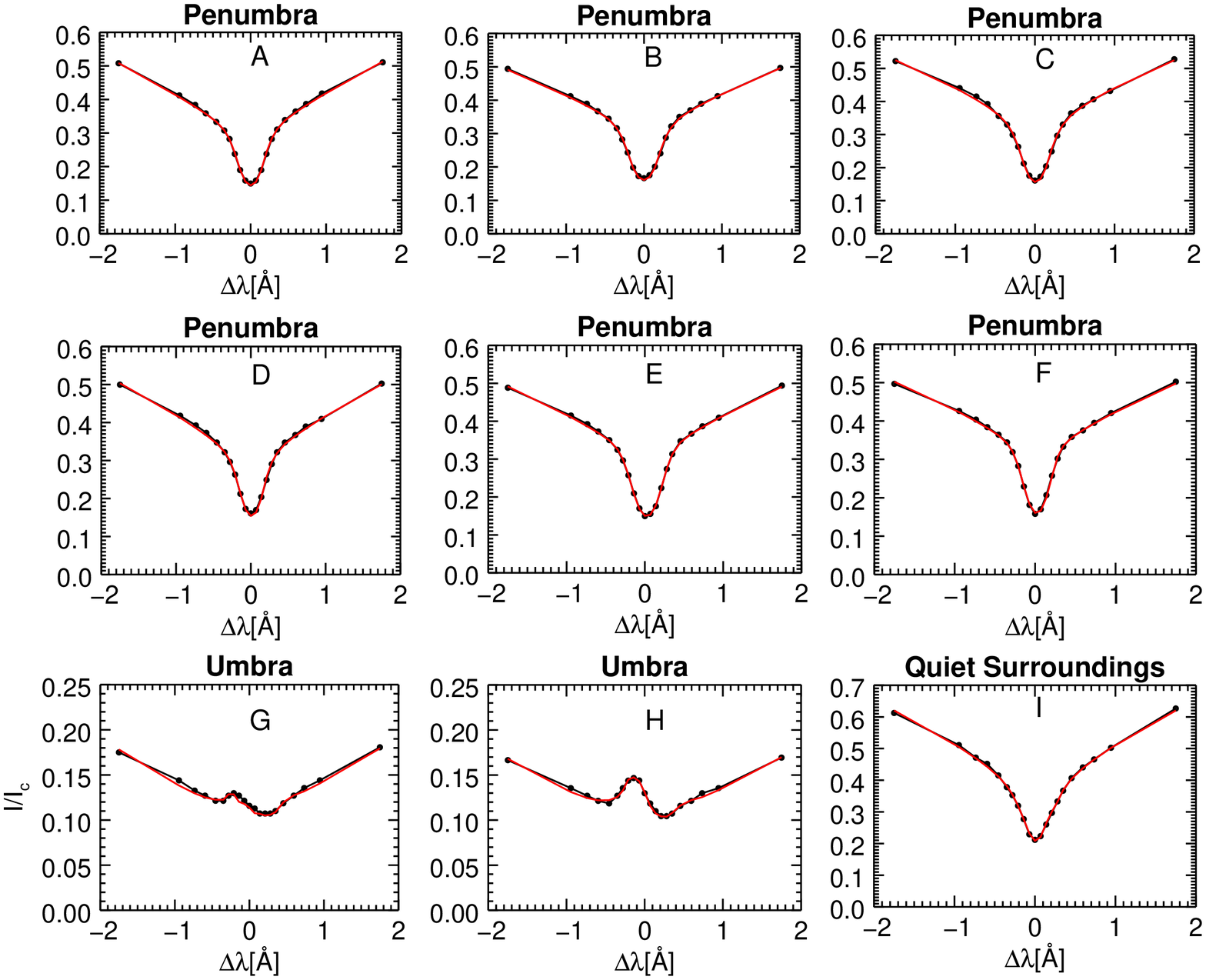}
   \caption{Comparison of the observed (black dots) and their corresponding synthetic (red) \ion{Ca}{ii}~8542~\AA\ Stokes I profiles for the pixels shown in Fig.~\ref{figure:rh_vs_iris_spectra} above. The fits show a very good agreement with the observed spectra that further reinforces the reliability of the model atmospheres used to synthesize the \ion{Mg}{ii}~h\&k spectra. }
              \label{figure:nicole_inv_observed}%
    \end{figure*}

A closer look at Fig.~\ref{figure:rh_vs_iris_spectra} reveals that for most of the pixels across the different features in the FOV, the synthetic spectra corresponding to the \ion{Mg}{ii} triplet, as well as  the photospheric spectral regions between k$_\mathrm{2r}$ and h$_\mathrm{2v}$ peaks along with the blended lines, match well with the IRIS observations. 
Further, the \ion{Mg}{ii} triplet lines bear structural resemblance to the Ca atom where the infrared triplets (8498, 8542, and 8662 \AA) are allocated to the higher levels of the corresponding h \& k lines and are known to sample mid chromospheric regions \citep{2015ApJ...806...14P}. A reasonable match of the synthetic spectra with the IRIS observations in these wavelength regions also builds confidence to our approach. 

More interestingly, the k$_\mathrm{1}$, k$_\mathrm{2}$ and h$_\mathrm{1}$, h$_\mathrm{2}$ line features from the observations are also well reproduced by the synthesis, including the peak separations. We verify that this is indeed the case for a vast majority of the penumbra of the sunspot. Six selected profiles corresponding to different locations in the penumbra are shown in Figs.~\ref{figure:rh_vs_iris_spectra} (A)-(F). The sub Figs.~\ref{figure:rh_vs_iris_spectra}-(A) and (E) in particular, show the spectra derived from the hot and cool penumbral model atmospheres and respective observations, as indicated in the bottom two rows of Fig.~\ref{figure:stokes_and_models}. We also note from Fig.~\ref{figure:stokes_and_models}, that the hot penumbra has a relatively weak horizontal or perpendicular (given by $B_\mathrm{HOR} =\sqrt{B_{x}^2+B_{y}^2}$; $B_{x}$ and $B_{y}$ being the fields in the horizontal x and y plane respectively) and LOS magnetic field compared to the cool penumbra. This is the case for a vast majority of different hot penumbral pixels in the FOV, where the cool penumbral magnetic field is greater than its hot counterpart.

Each sub figure in Fig.~\ref{figure:rh_vs_iris_spectra} has an inset window that shows the observed and the synthetic spectra for the \ion{Mg}{ii}~k region, for the sake of clarity and better visualization. The sufficiently good quality match (except for the line core) between the observed and the synthetic spectra indicates that the stratification of the different physical parameters are indeed good enough to capture the features of two very different spectral line formed over a wide range in the solar chromosphere. \citet{Leenaarts2013b} describes the $\tau$ =1 distribution of the k$_\mathrm{1}$, k$_\mathrm{2}$, h$_\mathrm{1}$, h$_\mathrm{2}$ peaks to be between 1.3 to 1.7 Mm that is of the same order of the peak sensitivity of the \ion{Ca}{ii}~8542 line core. Detailed analysis based on the contribution functions (discussed further in Sect.~\ref{sec:comparison_with_others} and Fig.~\ref{figure:temperature_stratification}) indicates that though the peak contribution of both the k$_\mathrm{2}$ wavelength and the 8542 line core is roughly around the same optical depth, the former has a wider spread towards lower optical depth and thereby samples atmospheric layers slightly higher than \ion{Ca}{ii}~8542. A successful reproduction of the \ion{Mg}{ii}~h\&k peak intensities, including their separation, demonstrates the strength of our models. It is known that the peak separation is indicative of the velocity variations from the mid to the upper chromosphere \citep{Leenaarts2013b}, and a satisfactory match between the synthetic and the observed profiles indicate that we are able to capture the intricate velocity variations with a significant level of accuracy. This suggests that these models can successfully explain the atmospheric properties above a sunspot penumbra up to the chromosphere. In this paper, we propose two models corresponding to a hot (Table~\ref{hot-table}) and a cool (Table~\ref{cool-table}) penumbra. To the best of our knowledge, these are the first chromospheric penumbral models put forward since the modelling attempts of \citet{Socas-Navarro2007models}, and the first to reproduce different chromospheric lines observed with different instruments (with one of them being a space-based slit-spectrograph and the other a ground-based Fabry-Perot instrument).

\subsubsection{Umbral Profiles}

The synthetic profiles obtained for the two umbral pixels as shown in Fig.~\ref{figure:rh_vs_iris_spectra} (G) and (H), show significant self-reversals compared to the observed ones. In almost all the pixels across the FOV we end up with a central depression in the synthetic line profiles whereas the observed ones indicate an emission core. \citet{Lites1982} also reported this behavior with more depressed central line cores in the umbra, however the reason behind this is still not fully clear.

Umbral profiles with strong asymmetries between the k$_\mathrm{2v}$, k$_\mathrm{2r}$, and h$_\mathrm{2v}$, h$_\mathrm{2r}$ peaks, 
similar to Fig.~\ref{figure:rh_vs_iris_spectra} (G) and (H), have been observed across multiple pixels in the FOV. The presence of a strong emission in the blue and a corresponding suppressed red k$_\mathrm{2}$ peak is a possible indication of strong gradient in the down-flows in the upper chromosphere and TR. Such strong down-flows were part of the models that best reproduced some umbral flashes in \citet{Vasco2017-Flashes}. In fact \citet{Lites1982} also introduced a strong downdraft of about 40~\kms\ on top of their atmosphere in order to take into account this asymmetry in their umbral profiles. We discuss this further in Sect. \ref{sec:comparison_with_others}.

Except for the self-reversal, the synthetic spectra in Fig.~\ref{figure:rh_vs_iris_spectra}~(G) seems to capture the shape of the observed line profile over the entire spectral range, including the peaks and  exterior slopes of h$_{2}$ and k$_{2}$ (as an envelope). Unlike the penumbra, the signal level between k$_\mathrm{2r}$ and h$_\mathrm{2v}$ is negligible and therefore it is difficult to comment on the match between the observed and the synthetic spectra in this wavelength range; nevertheless based on the reasonable matches in the k$_\mathrm{2}$ and h$_{2}$ peaks we feel confident of the strong down-flow scenario in our model atmosphere. This is further clear from the velocity stratification shown in Fig.~\ref{figure:stokes_and_models}~(third column), where the \verb|NICOLE| inversions reveal a strongly positive LOS velocity of up to 10 \kms with decreasing optical depth. Furthermore, the Stokes parameters corresponding to the umbral pixel in Fig.~\ref{figure:stokes_and_models} and Fig.~\ref{figure:nicole_inv_observed} (G) show a typical flash-like behavior with an emission feature. The model atmosphere corresponding to such an umbral flash has been proposed in Table~\ref{flash-table}.

Supersonic downflows, of the order of 100 \kms, have been observed on a number of occasions in the TR above sunspot umbrae in the recent past \citep{2014ApJ...789L..42K,2015A&A...582A.116S,Chitta2016,2018ApJ...859..158S}. Some of these studies \citep[such as][]{Chitta2016,2018ApJ...859..158S} reported signtaures of strong downflows in the chromospheric \ion{Mg}{ii}~h\&k lines, that originate in the TR and were found to be associated with coronal loops and sometimes in a coronal rain. \citet{Chitta2016} argue that the high speed supersonic downflows in the corona would have to undergo a shock transition of subsonic speeds, lower in the atmosphere and affect the chromospheric lines. 

\begin{figure}[htb!]
   \centering
   \includegraphics[width=\hsize]{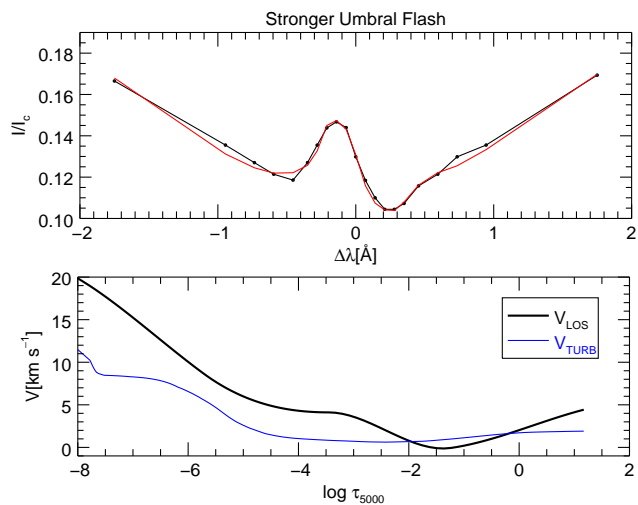}
   \caption{Top panel: Umbral flash intensity profile corresponding to a region of strong down-flow as shown in figure \ref{figure:rh_vs_iris_spectra} (H). Bottom panel: The corresponding stratification of LOS and microturbulent velocity. We clearly see that the LOS velocity is higher in this case. }
              \label{figure:supersonicflash}%
    \end{figure}

Fig.~\ref{figure:rh_vs_iris_spectra} (H) (also Fig.~\ref{figure:nicole_inv_observed} (H)), indicates the presence of another strong super-sonic down-flow in an umbral flash observed in the \ion{Mg}{ii}~h\&k spectral line (and in the \ion{Ca}{ii}~8542 \AA\ spectra), that is mainly chromospheric. Line-of-sight velocities up to +20~\kms\ with a strong gradient were found from the \ion{Ca}{ii}~8542 inversions for the corresponding pixel as shown in Fig.~\ref{figure:supersonicflash}. These velocities are stronger than the one shown in Fig. \ref{figure:stokes_and_models}. The highest values of the velocity occur at the highest layers, towards the end of an up-trending slope, where \ion{Ca}{ii}~8542 is no longer well constrained. But the synthetic \ion{Mg}{ii} spectra show a direct consequence of this strong down-flow by an enhanced blue-red peak asymmetry, that is stronger than what was found in Fig.~\ref{figure:rh_vs_iris_spectra} (G). The significantly higher Doppler-shifted IRIS \ion{Mg}{ii} spectra indicates a possibility of even stronger down-flows that are not captured by the \ion{Ca}{ii}~8542 inversions. Statistical investigations by \citet{2018ApJ...859..158S} report the detection of down-flows of the order of at least 40~\kms\ in both the penumbra and umbra of different sunspots observed by IRIS, between September 2013 and April 2015, across multiple chromospheric and TR lines. Though they do not indicate the presence of umbral flashes in their observations, it is possible that sometimes these downflows in the umbra maybe associated with a flash, as we report in this paper. With such a high gradient in the LOS velocity, it would be enough to cause a significant redshift of the k$_{3}$ and h$_{3}$ line core and cause the opacity to shift in such a way that it causes a stronger emission in the blue and a suppressed emission peak in the red. Our model atmospheres however, do not include velocities much higher than 20~\kms, even for the most extreme models and at heights where the observations do not allow a good constraint; this could be a possible reason why the synthetic umbral flash \ion{Mg}{ii} profiles do not reproduce the shift that is present in the observed profiles.

Since its discovery, umbral flashes have been understood as a manifestation of up-flows \citep{1969SoPh....7..351B}. Recent semi-empirical investigations of umbral flashes by \citet{2000ApJ...544.1141S,2000Sci...288.1396S,Jaime2013_Flashes} and \citet{2018A&A...619A..63J} support such interpretation by obtaining, via inversions, purely up-flowing atmospheres that reproduce the observed umbral flash profiles post non-LTE CRD radiative transfer of the Ca II IR lines. \citet{Bard_and_Carlsson_2010} modelled umbral flashes in the \ion{Ca}{ii}~H\&K lines using hydrodynamic simulations followed by non-LTE CRD synthesis. They concluded that flashes are a result of acoustic waves generated in the photosphere which steepen into a shock in the chromosphere. Similarly, \citet{Felipe_2014} synthesized full Stokes spectropolarimetric profiles in the Ca~II~IR of an umbral flash generated by a numerical simulation. In both studies the contribution functions to the wavelength of the flashed blue emission peaks are highest at atmospheric heights that feature up-flowing atmospheres. This, together with the majority of the semi-empirical work mentioned and the success of the simulation synthetics in reproducing the properties of umbral flashes, including spectral evolution in time, lends strong support for a model of umbral flashes that is upflowing. However, the highest opacity and thus highest formation height is at the line minimum, which is captured in the observations (and thus the inversions). Both numerical models include strong downflows in the upper layers of the solar atmosphere, immediately above the weaker upflows. Further, the flash formation of \citet{Bard_and_Carlsson_2010} is remarkably similar to that of bright grain formation \citep[see][]{1997ApJ...481..500C} and, in the latter, it is made unambiguously clear that the opacity shift produced by the strong downflows is critical for the formation of the strong blue peak itself. This is similar to the opacity effect described by \cite{1984mrt..book..173S} and also important for the line formation of magnetic-bubbles observed in flux emergence regions \citep{2015ApJ...810..145D}. As far as we are aware, there are no simulations of umbral flashes with purely up-flowing atmospheres and the downflows are always stronger than the upflows. The agreement of previous semi-empirical results in the literature means that we should still regard umbral flashes as likely occurring in, or leading to, upflowing atmospheres. We however find strongly down-flowing models, similar to those of \citet{Vasco2017-Flashes}, but reproducing both the observed \ion{Ca}{ii} and \ion{Mg}{ii} flash profiles. In addition, unlike \citet{Vasco2017-Flashes} and the other semi-empirical modelling studies, we do not find atmospheres that reproduce umbral flashes with just upflows; this, coupled with the observed asymmetries in the k$_{2}$ and h$_{2}$ peaks, lead us to believe that umbral flashes, even at the maximum intensity stage when their blue emission peak is strongest, can be formed under downflowing conditions in the Sun. Finally, such downflows might modulate the location at which up-ward propagating waves steepen into shocks, similarly to one of the scenarios described in \cite{2017A&A...605A..14N} to explain small umbral brightenings.

\subsubsection{Quiet Surroundings: Located at the edge of the FOV}
\begin{figure}
   \centering
   \includegraphics[width=\hsize,scale=0.8]{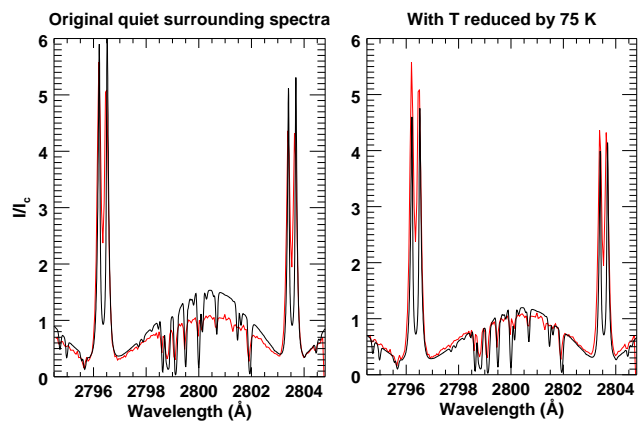}
      \caption{Comparison of the quiet surrounding \ion{Mg}{ii}~h\&k spectrum synthesized by RH1.5D (black) and IRIS observation (red). Left Panel: Synthetic spectrum from original model atmosphere. Right panel: Re-synthesized spectra with the temperature manually reduced by 75~K across all layers as described in the text. 
              }
         \label{figure:corrected_QS_profile}
   \end{figure}

The quiet surroundings are basically defined to be the regions around the sunspot, but close to it, where there is not any magnetic activity going on with time. We chose these regions based on visual inspection with \verb|CRISPEX| \citep{VissersCRISPEX2012}, an IDL widget-based tool designed for effective visualization. Unlike the penumbra and the umbra, the quieter areas on the inverted FOV (marked with a square in Fig.~\ref{fig:context}), have mismatches in the extended wings of the \ion{Mg}{ii} spectrum as seen in the left panel of Fig.~\ref{figure:corrected_QS_profile}, over the entire spectral range of interest. Based on these differences between the observed and the synthetic Mg spectra, it is our contention that the discrepancy in the far wings  of the profiles could be due to a small difference in the temperature stratification compared to what was obtained from the inversions. Since the synthetic profiles have higher intensities in the far wings than the observations, we re-synthesized the \ion{Mg}{ii} spectra with an atmosphere where the temperature was reduced by 75~K as a whole. As expected, this resulted in acceptable matches between the observed and the synthetic profiles (Fig.~\ref{figure:corrected_QS_profile}: right panel) which otherwise were distinctly different in the original synthesis. The spectra in Fig.~\ref{figure:rh_vs_iris_spectra} (I) (same as in the right panel of Fig. \ref{figure:corrected_QS_profile}) shows the synthetic spectra after the manual adjustment described above. The synthetic profiles from the adjusted atmospheres provide a good match with the observed spectra in the line wings, k$_{2}$ and h$_{2}$ features, and the far wings.

Synthesizing the \ion{Ca}{ii}~8542~spectra for the modified atmosphere, allowing for self-consistent changes in the hydrodynamic parameters via \verb|NICOLE's| equation of state, led to a profile that was no longer a perfect fit to the observations, with the synthetic profile being lower in intensity at all wavelengths. In this regard, we describe the model atmosphere for the quiet surrounding in Table \ref{Quieter-table} below, that corresponds to the original \verb|NICOLE| inversions (i.e. without the 75~K change). Therefore, unlike the case of the penumbra, we find a situation where multiple lines cannot successfully be reproduced by our approach. In such situations the best approach is likely to attempt multi-line PRD inversions, recently made available with the STiC code \citep{2019A&A...623A..74D}. Nevertheless, the quiet atmosphere, with the proposed temperature adjustment, could prove to be an excellent starting atmosphere for inversions.

\subsection{Magnetic field stratification in the model atmospheres}
The semi-empirical models shown in Tables \ref{hot-table}, \ref{cool-table}, and \ref{flash-table} have a constant magnetic field with height. The Mg spectra were synthesized by averaging the value of $B_\mathrm{LOS}$ and $B_\mathrm{HOR}$ between $\log\tau = [-6,-2]$ for the penumbra and only $B_\mathrm{LOS}$ for the umbra, where the sensitivity of the \ion{Ca}{ii}~8542~\AA\ would be maximum \citep{Carlos2016}. This approach ensures the $B_\mathrm{LOS}$ does not change change polarity with height as obtained from the \verb|NICOLE| inversions (Fig.~\ref{figure:stokes_and_models}) and provide a more consistent picture of the magnetic field. $B_\mathrm{HOR}$ for the umbra was set to zero for all heights and the model descrining the quiet surroundings does not include magnetic field.

\subsection{Synthetic vs observed spectroheliograms}

\label{sect:syn_vs_obs}

\begin{figure}
   \centering
   \includegraphics[width=\hsize]{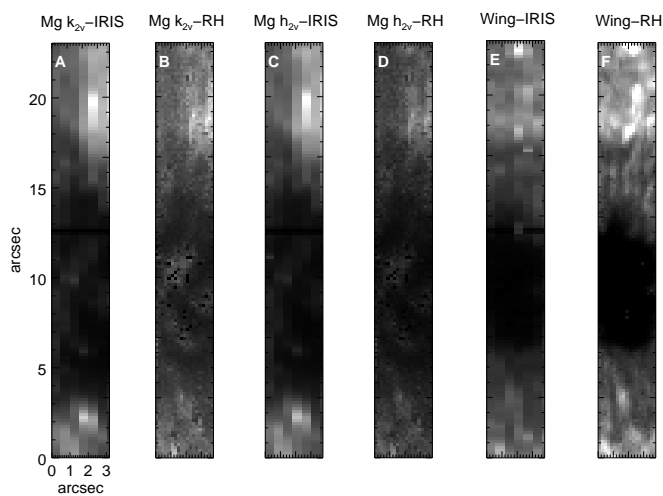}
      \caption{Comparison of the images
obtained from the RH synthesis and images composed from the IRIS slit scan
observations for the inverted FOV and for a narrow range of wavelengths: (A) Mg k$_\mathrm{2v}$ IRIS image, (B) Mg k$_\mathrm{2v}$ image obtained from RH1.5D synthesis, (C) Mg h$_\mathrm{2v}$ IRIS image, (D) Mg h$_\mathrm{2v}$ image obtained from RH1.5D synthesis, (E) \ion{Mg}{ii}~k wing image at 2794.7 \AA\ from IRIS, and (F) \ion{Mg}{ii}~k wing image at 2794.7 \AA\, from RH1.5D synthesis. The spectra were integrated over a definite wavelength window as described in Sect.~\ref{sect:syn_vs_obs}. Each pair of images was scaled between a common intensity range for efficient comparisons. }
         \label{figure:Synthetic_vs_realimage}
   \end{figure}

 We generate spectroheliograms from both the IRIS observations and RH1.5D synthesis and compare them in Fig.~\ref{figure:Synthetic_vs_realimage}. They were integrated over 0.6~\AA\ windows around 2796.2~\AA\ (k$_\mathrm{2v}$), 2803.33~\AA\ (h$_\mathrm{2v}$), and 2794.7~\AA\ (wing). Because IRIS was observing an 8-step sparse raster with 1\arcsec\ steps of the 0\farcs33 wide spectrograph slit, there are gaps with missing data in the $x$ direction. For an easier visual comparison of the raster maps, we widened the data to fill the gaps. 
This further emphasizes the difference in resolution between IRIS and the synthetic profiles (derived from SST observations). Nevertheless, the synthetic images agree well with the IRIS observations, highlighting the reliability of the semi-empirical models used to generate them.

\subsection{The deeper k$_{3}$ and h$_{3}$ line cores}

The most distinct difference between the observed and the synthetic spectra is the fairly cooler k$_{3}$ and h$_{3}$ line cores in the synthetic profiles for almost all the pixels in the FOV. This is clear from Fig.~\ref{figure:rh_vs_iris_spectra}. It can be explained by the fact that the \ion{Mg}{ii}~h and k line cores form higher than the \ion{Ca}~8542 line core. Therefore, the sensitivity of the inversions is fairly low at such heights, impacting the validity of the resulting models in those locations. Studies by \citet{Leenaarts2013a,Leenaarts2013b} and \citet{2015ApJ...806...14P} for example, have shown that the peak sensitivity of the \ion{Mg}{ii} line core is typically less than 200 km below the TR (or around heights between 2 to 3 Mm in their model) . The inverted model atmospheres obtained from \ion{Ca}~8542 are certainly not sensitive to those heights. 

Further, as shown in \citet{Carlsson_2015_plage}, single-peaked or flat-topped profiles in Mg can possibly be explained by shifting the TR to a higher column mass. Though this approach fills up the k$_{3}$ minimum in the synthetic profiles to some extent, their best fit model still has some central reversal unlike the observations. Nevertheless, this mechanism can indeed provide a good starting point in investigating the single peaked profiles. In our analysis, the initial guess model (FAL-C) has the TR at a lower column mass. NICOLE does not change or attempt to fit the upper most point of the reference hydro-dynamical variable (in our case electron pressure) which makes it hard for the TR to shift greatly in column mass. More recent codes such as STiC \citep{2019A&A...623A..74D} have a mechanism to allow changes to the boundary condition and thus may be able to address the core of such profiles in the future.
Nonetheless, the 1.5D approach works well in reproducing the h\&k line profiles outside the very cores, and lets us select a number of interesting semi-empirical candidate atmospheric models, which is the main intent of this paper.

\subsection{Where do our models stand in comparison with earlier approaches?}
\label{sec:comparison_with_others}
There have been some attempts in the past to model sunspot atmospheres but they have been mostly restricted to the photosphere \citep{Iniesta94,luc2002,Fontenla2006}. Out of these, the models from \citet{Iniesta94} and \citet{luc2002} were based on spectral inversions whereas \citet{Fontenla2006} used "radiation-effective" forward modeling approach to construct the semi-empirical model atmospheres. \citet{Fontenla2009} (also based on forward modeling approach) was an improvement over the former \citep{Fontenla2006} model atmospheres due to the fact that it was the first time that they introduced upper-chromospheric layers in these models for the quiet Sun and active region features and they further computed the spectra from these semi-empirical model atmospheres to compare with the actual observations. However, comparisons with the observed upper chromospheric spectra of \ion{Ca}{ii}~H\&K and \ion{Mg}{ii}~h\&k \citep[see also][]{Fontenla2011JGR} revealed that there was a significant mismatch between the two, thereby highlighting the shortcomings in their models. One of the major reasons was the lack of PRD in their radiative transfer calculations. We, on the other hand took PRD into consideration while computing the spectra from our model atmospheres presented in this paper and ended up with reasonable matches with the observed IRIS \ion{Mg}{ii} spectra for a number of pixels over a sunspot. This enhanced the strength and the applicability of our models. 

\citet{Lites1982} proposed an umbral model for the upper chromopshere and TR based on the observations in the Lyman~$\alpha$, \ion{Ca}{ii}~H\&K and \ion{Mg}{ii}~h\&k spectra. Their models did not account for the magnetic field, but they performed radiative transfer calculations under PRD. Assuming a strong down-flow of about 40~\kms\ on top of their model atmosphere, they were able to reproduce the observed asymmetries in the umbral line profiles. They further showed that the profiles are sensitive to temperature and both the LOS and microturbulent velocity, which we also find in our investigation. However, neither their observations included a flash, nor could they reproduce the single emission profiles that is mostly observed in the umbra. Nevertheless, it remains one of the earliest known attempts to obtain an umbral atmosphere that intends to describe the blue-red peak asymmetry in the Mg line profile. \citet{Maltby1986}, also provided semi-empirical model atmospheres for the dark umbral cores and their variation with the solar cycle that remains, along with \citet{Lites1982}, one of the earliest known models for the chromosphere of the umbra. We however, propose an umbral model in this paper that has a supersonic downflow associated with it, similar to \citet{Vasco2017-Flashes}, but supported further by \ion{Mg}{ii} observations.

\begin{figure*}
   \centering
   \includegraphics[scale =0.8,angle=90]{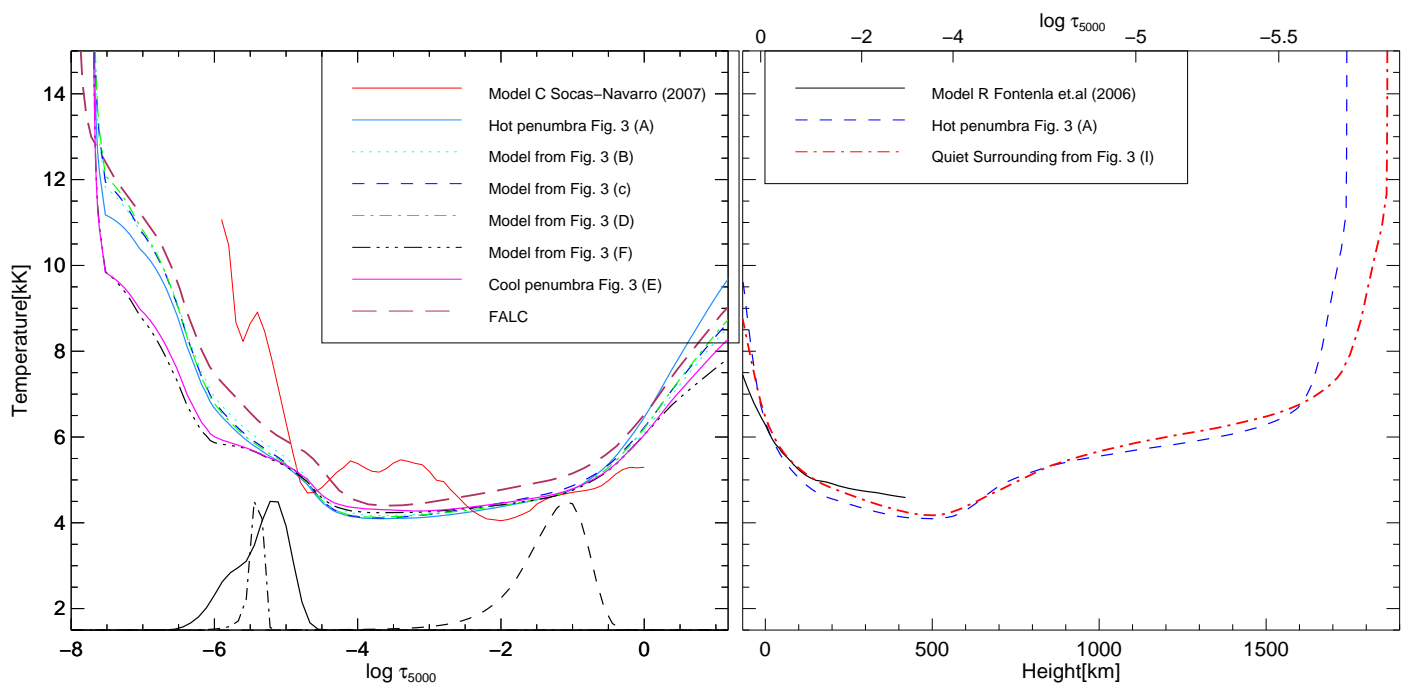}
   \caption{Left panel: Temperature stratification of six different models corresponding to the six different pixels in the penumbra of the sunspot (shown in Fig.~\ref{figure:rh_vs_iris_spectra}) as a function of optical depth. The model with the lowest uncertainty from \citet{Socas-Navarro2007models}, the work presenting the only two penumbral chromospheric models preceding this paper, is also shown for the sake of comparison. Contribution functions computed from the cool penumbral model for the \ion{Mg}{ii}~k$_\mathrm{2v}$, \ion{Ca}{ii}~8542 core and \ion{Ca}{ii} wing-averaged over the wavelength range between 8541.15 \AA\ and 8542.28 \AA, are plotted (in arbitrary units) with solid, dash-dotted and dashed lines respectively. Right panel: Temperature stratification as a function of geometric and optical depth of our hot penumbral and quiet surrounding model compared to the purely photospheric model-R of \citet{Fontenla2006}.}
    \label{figure:temperature_stratification}%
    \end{figure*}

The temperature stratification of the penumbral models presented in this paper not only compares reasonably well with \citet{Iniesta94}, \citet{luc2002}, and \citet{Fontenla2006} down to the photosphere, but it also extends higher up into the chromosphere. For the sake of completeness, we show also the temperature stratification of our models together with that of \citet{Socas-Navarro2007models} (Model-C) in the left-panel Fig.~\ref{figure:temperature_stratification} as a function of  $\log\tau_{5000}$. As far as we are aware, the latter is the only complete chromospheric model for the penumbra, including  hydrodynamical variables, preceding this work. More recently, \citet{2018A&A...619A..63J} modeled the temperature and $V_\mathrm{LOS}$ stratification, as a function of optical depth and up to $\log\tau = -5.5$, for the average quiescent and flash atmosphere of a sunspot, including both umbral flashes and running penumbral waves. Their models indicate that the temperature is hotter by about 0.2--0.5~kK in a running penumbral wave atmosphere between $\log\tau = -4.5$ and $-5$. They also find an increase in velocity of 1~km~s$^{-1}$ in LOS velocity when a wave is present in the penumbra.

The temperature variation of six different models in Fig.~\ref{figure:temperature_stratification}, corresponds to the six penumbral profiles in 
Fig.~\ref{figure:rh_vs_iris_spectra}. This also includes the cool and the hot penumbra. It is apparent that the cool models depart from their hot counterparts around $\log\tau = -5.5$, which is well within the sensitivity of both the \ion{Ca}{ii}~8542 core and \ion{Mg}{ii}~k$_\mathrm{2}$ peaks, with the sensitivity extending somewhat further upwards in the solar atmosphere. This is evident from the contribution function distributions that are also shown in the left panel of Fig.~\ref{figure:temperature_stratification}. This departure is stronger than other variations at lower heights and occurs at optical depths not modelled in any previous work. Being close to the limit of detection, it may be that this departure can only be identified with the benefit of a slight inclination in the LOS (that being $\mu$=0.745 in this case), which slightly increases the length of our atmosphere that is under upper chromospheric conditions. 
From the figure it is also clear that the height dependence of our derived models is rather gradual and smooth, unlike the Model-C by \citet{Socas-Navarro2007models} that has large uncertainties beyond $\log\tau = -5$. This is perhaps due to the earlier lack of adequate data to properly constrain the inversions. All the penumbral models shown in the left panel of Fig. \ref{figure:temperature_stratification} have their TR shifted deeper, as measured in the $\log\tau_{5000}$ scale, compared to the FAL-C atmosphere. For the sake of completeness, we also show the temperature stratification of our hot penumbra and the quiet surrounding model, along with a purely photospheric penumbral model from \citet{Fontenla2006} in the right panel of Fig.~\ref{figure:temperature_stratification}

The approaches by various authors described above, yielded several atmospheres over the past years. However, as discussed, a model complete with hydro-dynamical variables, that is effectively constrained from the photosphere to the chromosphere, is found to be lacking for sunspots, especially for the penumbra. We hope that the models proposed in this paper will fill in that void and be used as useful reference or starting atmospheres to describe the penumbra of sunspots in the future. 

\section{Conclusions}

\label{sec:conclusion}
The goal of this work is to obtain semi-empirical model atmospheres for a sunspot that work well for both the \ion{Ca}{ii} and \ion{Mg}{ii} spectra and are well constrained from the photosphere to the chromosphere. We chose to invert the chromospheric \ion{Ca}{ii}~8542 \AA\, full Stokes profiles observed with the CRISP instrument at the SST, with the \verb|NICOLE| inversion code to obtain the model atmospheres. To test the strength of the models we synthesized \ion{Mg}{ii}~h\&k line profiles with the help of RH1.5D radiative transfer code with the chromospheric models as inputs.

The synthetic spectra were compared with co-temporal and co-spatial IRIS observations over the full FOV. Detailed comparisons showed that our models reasonably reproduced the entire observed spectra of the penumbra, an umbral flash and a quiet surrounding region of the sunspot, with the exception of the line cores. Since the \ion{Mg}{ii} lines are sensitive to the upper chromospheric conditions, this comparison served as a consistency check for the models that have been obtained purely from \ion{Ca}{ii}~8542 inversions. It also indicated how well our models were constrained across multiple heights in the solar atmosphere. Based on these matches, we propose two penumbral models (hot penumbra and cool penumbra) in Tables \ref{hot-table} and \ref{cool-table}, respectively, reflecting a departure in temperature observed in the upper chromosphere. 

Comparing the temperatures we obtain over the formation range of the \ion{Mg}{ii}~h\&k peaks with those of \cite{2019arXiv190105763J} for Atacama Large Millimeter/submillimeter Array's \citep[ALMA:][]{2009ALMA} band 6 (1.3 mm), we find a good match for the penumbral case (6 to 7~kK in both works). This suggests that ALMA's band 6 forms at the same height as the peaks of the k$_{2}$ and h$_{2}$ when non-LTE is accounted for, or slightly higher, likely in the h$_{3}$ and k$_{3}$ range, once one considers the difference in the observing angle.

The umbral model as described in Table \ref{flash-table} reproduces the profiles of a typical umbral flash, but with supersonic down-flows reaching a maximum of 12 \kms\ in the upper layers of the solar atmosphere. We found that velocities such as these are essential to account for the strong blue-red peak asymmetries as observed in \ion{Mg}{ii}~h$_\mathrm{2v}$~\&~h$_\mathrm{2r}$ and also k$_\mathrm{2v}$ and k$_\mathrm{2r}$ umbral spectra. Based on recently published works, we also speculate that stronger down-flows of the order of 30--40 \kms\ could be present in the upper chromosphere and the TR in strong umbral flashes that may have a relation to coronal loops rooted deep into the umbra.

Furthermore, we also find that a slightly modified atmosphere, 75~K colder than the one obtained from inversions, leads to a much better match in the far wings for the \ion{Mg}{ii} quiet-sun like profiles which otherwise has a distinct mismatch. This behavior shows the strong dependence of the majority of \ion{Mg}{ii} profiles on temperature. We describe the quiet surrounding atmosphere in Table~\ref{Quieter-table}.

In all cases, we obtain models that reproduce all spectral features with the exception of the \ion{Mg}{ii}~k$_{3}$ and h$_{3}$ line cores. Our procedure is limited by \ion{Ca}{ii}~8542 spectra not being sensitive to the height of formation of k$_{3}$ and h$_{3}$. We anticipated that large changes at heights where \ion{Ca}{ii}~8542 is not very sensitive would, as part of the single-line fitting procedure, lead to some selected atmospheres that would approximately reproduce the k$_{3}$ and h$_{3}$ line cores. However, the bias at heights where sensitivity is low seems to go in the direction of producing colder atmospheres. This, along with the fact that NICOLE cannot move the TR to a higher column mass, may have prevented the reproduction of line-cores. Further, when dealing with active regions it may be necessary to adjust the upper boundary condition in \verb|NICOLE|, namely the value of the hydrodynamic parameter used as a starting point for the hydrostatic equilibrium stratification, an adjustment that this method does not allow. These reasons must contribute to deeper and cooler cores in the synthesized spectra and can be taken into account in future investigations. Apart from these discrepancies, it is important to understand that we compare profiles obtained from two different but overlapping height ranges in the solar atmosphere, using data that were recorded from two different instruments. The matches between the observed and the synthetic profiles for both the \ion{Mg}{ii}~h\&k spectral region and the \ion{Ca}{ii}~8542 line, allow us to put forward semi-empirical models for the chromosphere with a higher degree of confidence than possible in previous works. Our results also highlight the consistency between the two independent radiative transfer codes.

\begin{acknowledgements}
We are indebted to Shahin Jafarzadeh and Ainar Drews for observing these data at the Swedish 1-m Solar Telescope and coordinating the IRIS co-observations. We are thankful to the anonymous referee for their comments and suggestions. We would like to thank Carlos Quintero Noda for his discussions. We would like to acknowledge the support by the Research Council of Norway, project number 250810, and through its Centers of Excellence scheme, project number 262622 and through grants of computing time from the Programme for Supercomputing. IRIS is a NASA small explorer mission developed and operated by LMSAL with mission operations executed at NASA Ames Research center. 
The Swedish 1-m Solar Telescope is operated on the island of La Palma by the Institute for Solar Physics of Stockholm University in the Spanish Observatorio del Roque de los Muchachos of the Instituto de Astrof\'{i}sica de Canarias. The Institute for Solar Physics is supported by a grant for research infrastructures of national importance from the Swedish Research Council (registration number 2017-00625).
\end{acknowledgements}

\bibliographystyle{aa} 

\begin{appendix}

\section{Tabulated form of the semi-empirical models}

In this appendix, we provide the model atmospheres for the cool (Table~\ref{cool-table}), and hot penumbra (Table~\ref{hot-table}), umbral flash (Table~\ref{flash-table}) and the quiet surrounding (Table~\ref{Quieter-table}) respectively, in a standard table format. These tables are also available electronically.
\onecolumn

\begin{table}[]
\centering
\caption{Semi-empirical model atmospheric parameters for the hot penumbra. The $B_\mathrm{LOS}$ is equal to 214.48 G and $B_\mathrm{HOR}$ is equal to 146.5 G for all heights. }
\label{hot-table}
\resizebox{\textwidth}{!}{%
\begin{tabular}{ccccccccc}
\hline
$\log\tau_{5000}$ & Height (km) & Temperature (K) & \begin{tabular}[c]{@{}c@{}}$V_\mathrm{LOS}$\\ (m s$^{-1})$\end{tabular} & \begin{tabular}[c]{@{}c@{}}$V_\mathrm{micro}$\\ (m s$^{-1})$\end{tabular} & \begin{tabular}[c]{@{}c@{}}$N_\mathrm{H_{tot}}$\\ (m$^{-3}$)\end{tabular} & \begin{tabular}[c]{@{}c@{}}$N_\mathrm{e}$\\ (m$^{-3})$\end{tabular} & \begin{tabular}[c]{@{}c@{}}$\rho$\\ (kg m$^{-3})$\end{tabular} \\
\midrule
-8.000 & 1744.147 & 88415.80 & 2801.12 & 11342.74  & 4.941e+05 & 8.212e+15 & 2.236e-11 \\
-7.767 & 1743.714 & 35886.21 & 2636.51 & 9921.28 & 1.351e+07 & 2.031e+16 & 5.531e-11 \\
-7.744 & 1743.103 & 28736.84 & 2619.55 & 9603.69  & 4.265e+07 & 2.537e+16 & 6.910e-11 \\
-7.716 & 1742.474 & 21448.55 & 2598.28 & 9230.50  & 3.115e+08 & 3.275e+16 & 9.379e-11 \\
-7.688 & 1741.958 & 16452.96 & 2576.94 & 8904.13  & 5.090e+09 & 4.267e+16 & 1.225e-10 \\
-7.670 & 1741.677 & 14563.79 & 2563.49 & 8750.85  & 2.616e+10 & 4.821e+16 & 1.384e-10 \\
-7.650 & 1741.375 & 13281.36 & 2547.98 & 8629.99  & 1.005e+11 & 5.288e+16 & 1.519e-10 \\
-7.619 & 1740.932 & 12438.50 & 2524.56 & 8527.31  & 2.811e+11 & 5.644e+16 & 1.624e-10 \\
-7.569 & 1740.181 & 11755.87 & 2485.82 & 8421.37  & 7.181e+11 & 5.955e+16 & 1.724e-10 \\
-7.493 & 1738.929 & 11153.57 & 2426.15 & 8320.97  & 1.792e+12 & 6.221e+16 & 1.829e-10 \\
-7.367 & 1736.366 & 11025.04 & 2326.12 & 8290.33  & 2.231e+12 & 6.314e+16 & 1.865e-10 \\
-7.181 & 1731.037 & 10720.53 & 2177.02 & 8224.20  & 3.797e+12 & 6.526e+16 & 1.951e-10 \\
-6.970 & 1721.974 & 10265.58 & 2011.09 & 8124.05  & 8.916e+12 & 6.922e+16 & 2.095e-10 \\
-6.749 & 1707.465 & 9715.99 & 1844.65 & 7988.06   & 2.802e+13 & 7.593e+16 & 2.312e-10 \\
-6.526 & 1686.144 & 8856.01 & 1690.74 & 7735.67   & 2.137e+14 & 8.882e+16 & 2.712e-10 \\
-6.309 & 1658.224 & 7795.31 & 1559.66 & 7304.40   & 4.410e+15 & 1.091e+17 & 3.418e-10 \\
-6.117 & 1624.854 & 7028.60 & 1462.92 & 6763.86   & 5.158e+16 & 1.145e+17 & 4.578e-10 \\
-5.941 & 1574.098 & 6557.72 & 1394.07 & 6236.97   & 1.917e+17 & 9.365e+16 & 7.011e-10 \\
-5.813 & 1513.174 & 6329.25 & 1356.97 & 5795.15   & 3.791e+17 & 8.313e+16 & 1.091e-09 \\
-5.718 & 1450.404 & 6170.02 & 1337.55 & 5436.53   & 6.566e+17 & 7.788e+16 & 1.707e-09 \\
-5.629 & 1377.409 & 6035.16 & 1322.88 & 5094.15   & 1.166e+18 & 7.684e+16 & 2.868e-09 \\
-5.545 & 1297.198 & 5917.21 & 1308.94 & 4743.88   & 2.138e+18 & 7.918e+16 & 5.097e-09 \\
-5.462 & 1213.885 & 5811.51 & 1295.43 & 4372.04   & 3.987e+18 & 8.403e+16 & 9.343e-09 \\
-5.378 & 1131.328 & 5721.67 & 1281.85 & 3966.16   & 7.399e+18 & 9.193e+16 & 1.718e-08 \\
-5.292 & 1053.377 & 5621.87 & 1268.10 & 3560.69   & 1.340e+19 & 9.640e+16 & 3.095e-08 \\
-5.204 & 981.679 & 5530.10 & 1253.97 & 3173.21    & 2.334e+19 & 1.007e+17 & 5.373e-08 \\
-5.108 & 915.365 & 5437.09 & 1238.75 & 2799.66    & 3.935e+19 & 1.027e+17 & 9.041e-08 \\
-5.005 & 853.993 & 5324.35 & 1222.61 & 2472.34    & 6.470e+19 & 9.790e+16 & 1.485e-07 \\
-4.898 & 797.356 & 5181.83 & 1206.00 & 2156.96    & 1.043e+20 & 8.519e+16 & 2.392e-07 \\
-4.788 & 744.541 & 5024.00 & 1189.04 & 1900.43    & 1.656e+20 & 7.182e+16 & 3.797e-07 \\
-4.674 & 695.532 & 4832.58 & 1171.63 & 1639.21    & 2.605e+20 & 5.969e+16 & 5.971e-07 \\
-4.561 & 653.802 & 4597.13 & 1154.41 & 1419.12    & 3.962e+20 & 5.659e+16 & 9.082e-07 \\
-4.447 & 622.481 & 4407.45 & 1137.25 & 1265.00    & 5.528e+20 & 6.721e+16 & 1.267e-06 \\
-4.328 & 597.786 & 4272.76 & 1119.65 & 1128.31    & 7.230e+20 & 8.015e+16 & 1.657e-06 \\
-4.195 & 574.639 & 4191.44 & 1099.93 & 1012.22    & 9.263e+20 & 9.434e+16 & 2.123e-06 \\
-4.034 & 549.903 & 4133.98 & 1076.47 & 911.45     & 1.203e+21 & 1.120e+17 & 2.758e-06 \\
-3.831 & 520.729 & 4101.84 & 1047.05 & 822.50     & 1.631e+21 & 1.389e+17 & 3.737e-06 \\
-3.561 & 483.854 & 4100.38 & 1008.59 & 727.04     & 2.376e+21 & 1.871e+17 & 5.447e-06 \\
-3.204 & 436.227 & 4114.42 & 954.39 & 628.79      & 3.813e+21 & 2.769e+17 & 8.827e-06 \\
-2.814 & 384.834 & 4168.77 & 878.54 & 525.51      & 6.337e+21 & 4.386e+17 & 1.467e-05 \\
-2.420 & 333.047 & 4256.59 & 785.46 & 478.56      & 1.039e+22 & 7.174e+17 & 2.408e-05 \\
-2.024 & 280.491 & 4367.86 & 677.34 & 533.85      & 1.689e+22 & 1.195e+18 & 3.912e-05 \\
-1.626 & 226.800 & 4493.61 & 556.53 & 627.63      & 2.729e+22 & 2.003e+18 & 6.323e-05 \\
-1.231 & 172.810 & 4632.97 & 426.63 & 842.48      & 4.348e+22 & 3.342e+18 & 1.008e-04 \\
-0.885 & 123.237 & 4878.05 & 339.81 & 1060.48     & 6.401e+22 & 5.723e+18 & 1.482e-04 \\
-0.585 & 77.334 & 5237.55 & 308.28 & 1254.70      & 8.724e+22 & 9.795e+18 & 2.018e-04 \\
-0.273 & 28.924 & 5819.11 & 309.23 & 1440.88      & 1.145e+23 & 2.281e+19 & 2.624e-04 \\
0.111 & -14.283 & 6756.06 & 344.39 & 1611.76      & 1.317e+23 & 1.227e+20 & 3.020e-04 \\
0.512 & -39.556 & 7928.25 & 403.81 & 1688.35      & 1.284e+23 & 7.189e+20 & 2.960e-04 \\
0.909 & -57.325 & 9022.77 & 466.76 & 1740.88      & 1.195e+23 & 2.526e+21 & 2.802e-04\\
\hline
\end{tabular}%
}
\end{table}
\begin{table}[]
\centering
\caption{Semi-empirical model atmospheric parameters for the cool penumbra. The $B_\mathrm{LOS}$ is equal to 1343.61 G and $B_\mathrm{HOR}$ is equal to 1561.73 G for all heights.}
\label{cool-table}
\resizebox{\textwidth}{!}{%
\begin{tabular}{ccccccccc}
\hline
$\log\tau_{5000}$ & Height (km) & Temperature (K) & \begin{tabular}[c]{@{}c@{}}$V_\mathrm{LOS}$\\ (m s$^{-1})$\end{tabular} & \begin{tabular}[c]{@{}c@{}}$V_\mathrm{micro}$\\ (m s$^{-1})$\end{tabular} & \begin{tabular}[c]{@{}c@{}}$N_\mathrm{H_{tot}}$\\ (m$^{-3})$\end{tabular} & \begin{tabular}[c]{@{}c@{}}$N_\mathrm{e}$\\ (m$^{-3})$\end{tabular} & \begin{tabular}[c]{@{}c@{}}$\rho$\\ (kg m$^{-3})$\end{tabular} \\
\midrule
-8.000 & 1725.286 & 87122.94 & 728.22 & 8454.59  & 5.154e+05 & 8.334e+15 & 2.269e-11 \\
-7.767 & 1724.869 & 34579.29 & 869.62 & 7415.96 & 1.614e+07 & 2.108e+16 & 5.740e-11 \\
-7.744 & 1724.282 & 27427.06 & 884.57 & 7183.91 & 5.596e+07 & 2.658e+16 & 7.243e-11 \\
-7.716 & 1723.682 & 20134.96 & 903.39 & 6911.22 & 5.423e+08 & 3.484e+16 & 9.995e-11 \\
-7.688 & 1723.202 & 15135.31 & 922.36 & 6672.75  & 1.533e+10 & 4.637e+16 & 1.331e-10 \\
-7.670 & 1722.946 & 13243.49 & 934.35 & 6560.75  & 1.048e+11 & 5.300e+16 & 1.522e-10 \\
-7.650 & 1722.673 & 11957.92 & 948.22 & 6472.44  & 5.361e+11 & 5.849e+16 & 1.689e-10 \\
-7.619 & 1722.273 & 11110.17 & 969.21 & 6397.42  & 1.901e+12 & 6.209e+16 & 1.828e-10 \\
-7.569 & 1721.591 & 10419.21 & 1004.11 & 6320.01 & 6.118e+12 & 6.512e+16 & 1.964e-10 \\
-7.493 & 1720.463 & 9803.81 & 1058.20 & 6246.65 & 1.970e+13 & 6.897e+16 & 2.098e-10 \\
-7.367 & 1718.176 & 9654.15 & 1149.56 & 6224.26 & 2.706e+13 & 7.048e+16 & 2.146e-10 \\
-7.181 & 1713.479 & 9325.36 & 1286.67 & 6175.94 & 5.588e+13 & 7.395e+16 & 2.254e-10 \\
-6.970 & 1705.646 & 8864.21 & 1439.43 & 6102.76 & 1.689e+14 & 7.968e+16 & 2.432e-10 \\
-6.749 & 1693.278 & 8347.34 & 1591.21 & 6003.40 & 6.782e+14 & 8.797e+16 & 2.695e-10 \\
-6.526 & 1675.111 & 7607.47 & 1728.20 & 5818.98 & 6.356e+15 & 1.002e+17 & 3.187e-10 \\
-6.309 & 1647.528 & 6754.56 & 1839.46 & 5503.86 & 8.152e+16 & 8.864e+16 & 4.439e-10 \\
-6.117 & 1591.181 & 6215.44 & 1914.82 & 5108.90 & 2.657e+17 & 5.464e+16 & 7.546e-10 \\
-5.941 & 1483.928 & 5963.00 & 1960.29 & 4723.91 & 6.653e+17 & 4.908e+16 & 1.650e-09 \\
-5.813 & 1387.891 & 5888.87 & 1977.50 & 4401.08 & 1.352e+18 & 5.880e+16 & 3.246e-09 \\
-5.718 & 1314.753 & 5835.11 & 1980.54 & 4139.05 & 2.317e+18 & 6.773e+16 & 5.479e-09 \\
-5.629 & 1247.096 & 5786.99 & 1977.29 & 3888.88  & 3.819e+18 & 7.743e+16 & 8.943e-09 \\
-5.545 & 1184.271 & 5737.75 & 1970.53 & 3632.94  & 6.093e+18 & 8.675e+16 & 1.418e-08 \\
-5.462 & 1125.302 & 5682.96 & 1960.77 & 3361.24  & 9.488e+18 & 9.457e+16 & 2.197e-08 \\
-5.378 & 1068.843 & 5626.97 & 1948.01 & 3064.68  & 1.456e+19 & 1.019e+17 & 3.360e-08 \\
-5.292 & 1014.794 & 5557.02 & 1932.29 & 2768.41  & 2.209e+19 & 1.051e+17 & 5.088e-08 \\
-5.204 & 963.200 & 5495.71 & 1913.47 & 2485.28  & 3.300e+19 & 1.098e+17 & 7.589e-08 \\
-5.108 & 913.011 & 5435.00 & 1890.51 & 2212.34  & 4.897e+19 & 1.144e+17 & 1.125e-07 \\
-5.005 & 864.302 & 5355.64 & 1863.51 & 1973.17  & 7.245e+19 & 1.132e+17 & 1.663e-07 \\
-4.898 & 817.129 & 5246.05 & 1833.40 & 1742.73  & 1.073e+20 & 1.033e+17 & 2.461e-07 \\
-4.788 & 771.021 & 5119.95 & 1800.76 & 1555.29  & 1.592e+20 & 9.090e+16 & 3.652e-07 \\
-4.674 & 725.797 & 4958.57 & 1765.94 & 1364.42  & 2.389e+20 & 7.572e+16 & 5.477e-07 \\
-4.561 & 683.237 & 4749.79 & 1730.84 & 1203.60  & 3.596e+20 & 6.418e+16 & 8.243e-07 \\
-4.447 & 647.910 & 4583.05 & 1696.00 & 1090.99  & 5.114e+20 & 6.980e+16 & 1.172e-06 \\
-4.328 & 619.563 & 4467.54 & 1661.16 & 991.12  & 6.814e+20 & 8.398e+16 & 1.562e-06 \\
-4.195 & 593.137 & 4401.64 & 1624.25 & 906.29  & 8.871e+20 & 1.034e+17 & 2.033e-06 \\
-4.034 & 565.973 & 4353.16 & 1585.00 & 832.66  & 1.162e+21 & 1.285e+17 & 2.664e-06 \\
-3.831 & 534.992 & 4321.13 & 1546.21 & 767.67  & 1.578e+21 & 1.648e+17 & 3.616e-06 \\
-3.561 & 496.785 & 4302.75 & 1520.46 & 697.92  & 2.295e+21 & 2.242e+17 & 5.260e-06 \\
-3.204 & 448.469 & 4276.44 & 1546.02 & 626.13  & 3.705e+21 & 3.254e+17 & 8.491e-06 \\
-2.814 & 396.318 & 4301.86 & 1645.70 & 550.66  & 6.076e+21 & 5.012e+17 & 1.407e-05 \\
-2.420 & 343.362 & 4367.05 & 1789.39 & 516.36  & 9.993e+21 & 7.975e+17 & 2.314e-05 \\
-2.024 & 289.566 & 4450.60 & 1942.17 & 556.76  & 1.636e+22 & 1.286e+18 & 3.789e-05 \\
-1.626 & 235.260 & 4551.44 & 2067.85 & 625.28 & 2.655e+22 & 2.091e+18 & 6.148e-05 \\
-1.231 & 180.041 & 4673.97 & 2129.63 & 782.27  & 4.271e+22 & 3.439e+18 & 9.894e-05 \\
-0.885 & 129.745 & 4869.13 & 2103.23 & 941.55 & 6.383e+22 & 5.666e+18 & 1.478e-04 \\
-0.585 & 84.055 & 5133.40 & 2021.40 & 1083.47  & 8.877e+22 & 9.164e+18 & 2.053e-04 \\
-0.273 & 34.528 & 5572.78 & 1892.92 & 1219.50  & 1.217e+23 & 1.731e+19 & 2.789e-04 \\
0.111 & -19.862 & 6267.29 & 1694.59 & 1344.36 & 1.593e+23 & 5.846e+19 & 3.651e-04 \\
0.512 & -59.310 & 7095.04 & 1468.08 & 1400.33  & 1.799e+23 & 2.529e+20 & 4.128e-04 \\
0.909 & -91.246 & 7829.55 & 1253.84 & 1438.71  & 1.938e+23 & 7.757e+20 & 4.461e-04\\
\hline
\end{tabular}%
}
\end{table}
\newpage
\begin{table}[]
\centering
\caption{Semi-empirical model atmospheric parameters for the umbral flash. The $B_\mathrm{LOS}$ is equal to 3313.25 G and $B_\mathrm{HOR}$ is zero, for all heights.}
\label{flash-table}
\resizebox{\textwidth}{!}{%
\begin{tabular}{ccccccccc}
\hline
$\log\tau_{5000}$ & Height (km) & Temperature (K) & \begin{tabular}[c]{@{}c@{}}$V_\mathrm{LOS}$\\ (m s$^{-1})$\end{tabular} & \begin{tabular}[c]{@{}c@{}}$V_\mathrm{micro}$\\ (m s$^{-1})$\end{tabular} & \begin{tabular}[c]{@{}c@{}}$N_\mathrm{H_{tot}}$\\ (m$^{-3})$\end{tabular} & \begin{tabular}[c]{@{}c@{}}$N_\mathrm{e}$\\ (m$^{-3})$\end{tabular} & \begin{tabular}[c]{@{}c@{}}$\rho$\\ (kg m$^{-3})$\end{tabular} \\
\midrule
-8.000 & 1654.248 & 90033.87 & 12111.70 & 10730.73 & 4.692e+05 & 8.064e+15 & 2.195e-11 \\
-7.767 & 1653.794 & 37810.17 & 11824.80 & 9326.73  & 1.059e+07 & 1.928e+16 & 5.249e-11 \\
-7.744 & 1653.146 & 30694.74 & 11796.83 & 9013.04  & 2.967e+07 & 2.376e+16 & 6.470e-11 \\
-7.716 & 1652.472 & 23449.14 & 11762.00 & 8644.43  & 1.561e+08 & 3.037e+16 & 8.541e-11 \\
-7.688 & 1651.901 & 18496.46 & 11727.24 & 8322.07  & 1.269e+09 & 3.796e+16 & 1.089e-10 \\
-7.670 & 1651.583 & 16634.27 & 11705.45 & 8170.68 & 4.424e+09 & 4.222e+16 & 1.212e-10 \\
-7.650 & 1651.236 & 15382.89 & 11680.35 & 8051.30 & 1.232e+10 & 4.567e+16 & 1.311e-10 \\
-7.619 & 1650.718 & 14586.59 & 11642.57 & 7949.88  & 2.566e+10 & 4.821e+16 & 1.384e-10 \\
-7.569 & 1649.831 & 13979.85 & 11580.16 & 7845.24  & 4.725e+10 & 5.038e+16 & 1.447e-10 \\
-7.493 & 1648.341 & 13490.22 & 11483.74 & 7746.08  & 8.039e+10 & 5.237e+16 & 1.504e-10 \\
-7.367 & 1645.260 & 13532.87 & 11319.21 & 7715.81  & 7.768e+10 & 5.254e+16 & 1.509e-10 \\
-7.181 & 1638.772 & 13420.09 & 11059.57 & 7650.49 & 9.051e+10 & 5.372e+16 & 1.543e-10 \\
-6.970 & 1627.627 & 13041.14 & 10734.64 & 7551.58 & 1.467e+11 & 5.661e+16 & 1.627e-10 \\
-6.749 & 1609.696 & 12340.98 & 10345.11 & 7417.26  & 3.821e+11 & 6.218e+16 & 1.790e-10 \\
-6.526 & 1583.246 & 11088.77 & 9889.73 & 7167.97  & 2.677e+12 & 7.240e+16 & 2.138e-10 \\
-6.309 & 1548.190 & 9517.37 & 9373.38 & 6741.99 & 5.820e+13 & 9.090e+16 & 2.770e-10 \\
-6.117 & 1510.208 & 8256.89 & 8844.09 & 6208.09  & 1.522e+15 & 1.179e+17 & 3.625e-10 \\
-5.941 & 1469.702 & 7353.24 & 8295.94 & 5687.68  & 2.718e+16 & 1.411e+17 & 4.870e-10 \\
-5.813 & 1431.003 & 6843.85 & 7850.20 & 5251.28  & 1.281e+17 & 1.307e+17 & 6.715e-10 \\
-5.718 & 1387.965 & 6508.08 & 7495.06 & 4897.07  & 2.934e+17 & 1.051e+17 & 9.613e-10 \\
-5.629 & 1325.962 & 6230.52 & 7156.98 & 4558.89 & 5.787e+17 & 8.335e+16 & 1.545e-09 \\
-5.545 & 1239.268 & 5991.23 & 6838.40 & 4212.93  & 1.196e+18 & 7.033e+16 & 2.918e-09 \\
-5.462 & 1131.062 & 5785.73 & 6532.82 & 3845.65  & 2.749e+18 & 6.540e+16 & 6.461e-09 \\
-5.378 & 1015.390 & 5616.16 & 6229.45 & 3444.76  & 6.648e+18 & 6.654e+16 & 1.540e-08 \\
-5.292 & 908.743 & 5439.70 & 5926.76 & 3044.27  & 1.537e+19 & 6.377e+16 & 3.537e-08 \\
-5.204 & 817.634 & 5267.88 & 5621.31 & 2661.55  & 3.228e+19 & 5.809e+16 & 7.411e-08 \\
-5.108 & 740.030 & 5088.01 & 5299.66 & 2292.59  & 6.245e+19 & 4.945e+16 & 1.432e-07 \\
-5.005 & 673.424 & 4883.30 & 4967.67 & 1969.29  & 1.136e+20 & 3.948e+16 & 2.604e-07 \\
-4.898 & 618.331 & 4647.59 & 4637.78 & 1657.78  & 1.933e+20 & 3.361e+16 & 4.431e-07 \\
-4.788 & 577.021 & 4397.33 & 4314.39 & 1404.41 & 2.992e+20 & 3.794e+16 & 6.858e-07 \\
-4.674 & 548.312 & 4115.43 & 3998.50 & 1146.39  & 4.235e+20 & 4.504e+16 & 9.708e-07 \\
-4.561 & 527.361 & 3796.38 & 3703.65 & 929.01  & 5.727e+20 & 4.050e+16 & 1.313e-06 \\
-4.447 & 508.172 & 3530.97 & 3429.59 & 776.78 & 7.587e+20 & 2.871e+16 & 1.756e-06 \\
-4.328 & 487.696 & 3328.18 & 3170.74 & 641.77  & 1.027e+21 & 1.994e+16 & 2.391e-06 \\
-4.195 & 464.538 & 3184.40 & 2910.10 & 527.11 & 1.422e+21 & 1.675e+16 & 3.362e-06 \\
-4.034 & 438.912 & 3077.41 & 2645.02 & 427.58 & 2.005e+21 & 1.705e+16 & 4.900e-06 \\
-3.831 & 410.283 & 3050.59 & 2389.47 & 339.72 & 2.889e+21 & 2.185e+16 & 7.313e-06 \\
-3.561 & 376.920 & 3148.82 & 2202.44 & 245.43 & 4.438e+21 & 3.698e+16 & 1.113e-05 \\
-3.204 & 337.707 & 3349.05 & 2049.13 & 148.39 & 7.139e+21 & 8.060e+16 & 1.737e-05 \\
-2.814 & 297.305 & 3473.04 & 1498.97 & 46.38  & 1.133e+22 & 1.496e+17 & 2.747e-05 \\
-2.420 & 254.185 & 3469.77 & 751.60 & 0.00  & 1.856e+22 & 2.122e+17 & 4.629e-05 \\
-2.024 & 208.486 & 3460.80 & 79.44 & 54.62  & 3.089e+22 & 3.074e+17 & 8.072e-05 \\
-1.626 & 161.803 & 3477.48 & -236.50 & 147.24  & 5.115e+22 & 4.749e+17 & 1.413e-04 \\
-1.231 & 115.809 & 3548.18 & 81.02 & 359.46  & 8.408e+22 & 7.999e+17 & 2.403e-04 \\
-0.885 & 77.213 & 3715.15 & 754.06 & 574.77  & 1.302e+23 & 1.484e+18 & 3.588e-04 \\
-0.585 & 46.587 & 3965.82 & 1384.56 & 766.61  & 1.816e+23 & 2.999e+18 & 4.695e-04 \\
-0.273 & 19.110 & 4399.76 & 2067.41 & 950.50 & 2.434e+23 & 7.884e+18 & 5.579e-04 \\
0.111 & -14.600 & 5101.98 & 2920.66 & 1119.28  & 2.825e+23 & 2.218e+19 & 6.475e-04 \\
0.512 & -60.880 & 5960.08 & 3787.71 & 1194.93  & 3.426e+23 & 5.926e+19 & 7.853e-04 \\
0.909 & -104.583 & 6729.88 & 4583.35 & 1246.82  & 4.047e+23 & 2.164e+20 & 9.281e-04\\
\hline
\end{tabular}%
}
\end{table}
\begin{table}[]
\centering
\caption{Semi-empirical model atmospheric parameters for the quiet surrounding. The magnetic field is absent in this case.}
\label{Quieter-table}
\resizebox{\textwidth}{!}{%
\begin{tabular}{ccccccccc}
\hline
$\log\tau_{5000}$ & Height (km) & Temperature (K) & \begin{tabular}[c]{@{}c@{}}$V_\mathrm{LOS}$\\ (m s$^{-1})$\end{tabular} & \begin{tabular}[c]{@{}c@{}}$V_\mathrm{micro}$\\ (m s$^{-1})$\end{tabular} & \begin{tabular}[c]{@{}c@{}}$N_\mathrm{H_{tot}}$ \\ (m$^{-3}$)\end{tabular} & \begin{tabular}[c]{@{}c@{}}$N_\mathrm{e}$\\ (m$^{-3})$\end{tabular} & \begin{tabular}[c]{@{}c@{}}$\rho$\\ (kg m$^{-3})$\end{tabular} \\
\midrule
-8.000 & 1865.676 & 89004.53 & 909.94  &  13303.88 &  4.848e+05 &  8.164e+15 &  2.221e-11 \\
-7.767 & 1865.237 & 36457.91 & 974.66  &  11882.42 &  1.258e+07 &  2.007e+16 &  5.453e-11 \\
-7.744 & 1864.616 & 29306.78 & 981.31  &  11564.83 &  3.833e+07 &  2.499e+16 &  6.789e-11 \\
-7.716 & 1863.974 & 22016.33 & 989.65  &  11191.64 &  2.520e+08 &  3.208e+16 &  9.134e-11 \\
-7.688 & 1863.442 & 17018.64 & 998.02  &  10865.27 &  3.305e+09 &  4.135e+16 &  1.181e-10 \\
-7.670 & 1863.150 & 15128.18 & 1003.30 &  10711.99 &  1.539e+10 &  4.657e+16 &  1.330e-10 \\
-7.650 & 1862.836 & 13844.31 & 1009.40 &  10591.13 &  5.401e+10 &  5.094e+16 &  1.455e-10 \\
-7.619 & 1862.372 & 12999.36 & 1018.61 &  10488.45 &  1.391e+11 &  5.432e+16 &  1.552e-10 \\
-7.569 & 1861.587 & 12313.58 & 1033.90 &  10382.51 &  3.327e+11 &  5.746e+16 &  1.644e-10 \\
-7.493 & 1860.284 & 11707.25 & 1057.57 &  10282.11 &  7.743e+11 &  6.030e+16 &  1.736e-10 \\
-7.367 & 1857.621 & 11574.88 & 1097.73 &  10251.47 &  9.553e+11 &  6.137e+16 &  1.771e-10 \\
-7.181 & 1852.086 & 11273.84 & 1159.15 &  10185.34 &  1.550e+12 &  6.349e+16 &  1.848e-10 \\
-6.970 & 1842.638 & 10841.06 & 1230.73 &  10085.19 &  3.262e+12 &  6.686e+16 &  1.977e-10 \\
-6.749 & 1827.328 & 10342.49 & 1307.69 &  9949.20  &  8.394e+12 &  7.210e+16 &  2.164e-10 \\
-6.526 & 1804.379 &  9546.44 & 1386.20 &  9696.81  &  4.520e+13 &  8.300e+16 &  2.509e-10 \\
-6.309 & 1773.802 &  8536.72 & 1462.71 &  9265.54  &  5.720e+14 &  1.022e+17 &  3.100e-10 \\
-6.117 & 1739.139 &  7802.42 & 1530.14 &  8725.00  &  5.496e+15 &  1.242e+17 &  3.865e-10 \\
-5.941 & 1699.475 &  7347.79 & 1590.71 &  8198.11  &  2.790e+16 &  1.432e+17 &  4.905e-10 \\
-5.813 & 1663.136 &  7121.77 & 1634.33 &  7756.29  &  6.714e+16 &  1.544e+17 &  6.073e-10 \\
-5.718 & 1630.667 &  6958.60 & 1666.05 &  7397.67  &  1.233e+17 &  1.587e+17 &  7.406e-10 \\
-5.629 & 1594.551 &  6815.30 & 1695.07 &  7055.29  &  2.094e+17 &  1.605e+17 &  9.332e-10 \\
-5.545 & 1553.203 &  6685.12 & 1722.39 &  6705.02  &  3.395e+17 &  1.608e+17 &  1.222e-09 \\
-5.462 & 1505.071 &  6563.95 & 1748.57 &  6333.18  &  5.427e+17 &  1.614e+17 &  1.678e-09 \\
-5.378 & 1448.031 &  6455.17 & 1774.54 &  5927.30  &  8.725e+17 &  1.652e+17 &  2.433e-09 \\
-5.292 & 1380.786 &  6332.76 & 1800.42 &  5521.83  &  1.484e+18 &  1.678e+17 &  3.831e-09 \\
-5.204 & 1301.957 &  6214.36 & 1826.52 &  5134.35  &  2.664e+18 &  1.752e+17 &  6.545e-09 \\
-5.108 & 1210.525 &  6089.27 & 1853.97 &  4760.80  &  5.191e+18 &  1.860e+17 &  1.235e-08 \\
-5.005 & 1109.752 &  5939.13 & 1882.26 &  4433.48  &  1.089e+19 &  1.916e+17 &  2.542e-08 \\
-4.898 & 1007.872 &  5755.19 & 1910.33 &  4118.10  &  2.334e+19 &  1.812e+17 &  5.392e-08 \\
-4.788 &  913.033 &  5552.68 & 1937.81 &  3861.57  &  4.890e+19 &  1.582e+17 &  1.124e-07 \\
-4.674 &  827.766 & 5313.74  & 1964.60 &  3600.35  &  9.813e+19 &  1.202e+17 &  2.252e-07 \\
-4.561 &  751.574 &  5030.49 & 1989.54 &  3380.26  &  1.916e+20 &  8.082e+16 &  4.393e-07 \\
-4.447 &  689.063 &  4793.38 & 2012.68 &  3226.14  &  3.431e+20 &  6.805e+16 &  7.864e-07 \\
-4.328 &  642.639 &  4611.21 & 2034.45 &  3089.45  &  5.372e+20 &  7.590e+16 &  1.231e-06 \\
-4.195 &  606.191 &  4479.43 & 2056.29 &  2973.36  &  7.729e+20 &  9.647e+16 &  1.772e-06 \\
-4.034 &  573.350 &  4368.14 & 2078.35 &  2872.60  &  1.084e+21 &  1.239e+17 &  2.484e-06 \\
-3.831 &  539.644 &  4284.63 & 2099.33 &  2783.64  &  1.533e+21 &  1.593e+17 &  3.513e-06 \\
-3.561 &  500.174 &  4251.23 & 2114.00 &  2688.19  &  2.269e+21 &  2.160e+17 &  5.200e-06 \\
-3.204 &  450.950 &  4280.42 & 2094.87 &  2589.94  &  3.660e+21 &  3.293e+17 &  8.390e-06 \\
-2.814 &  397.912 &  4366.82 & 1985.54 &  2486.65  &  5.956e+21 &  5.373e+17 &  1.379e-05 \\
-2.420 &  343.609 &  4485.59 & 1791.30 &  2439.70  &  9.685e+21 &  8.927e+17 &  2.242e-05 \\
-2.024 &  288.647 &  4628.22 & 1524.57 &  2495.00  &  1.557e+22 &  1.492e+18 &  3.605e-05 \\
-1.626 &  231.283 &  4791.76 & 1198.88 &  2588.77  &  2.507e+22 &  2.522e+18 &  5.804e-05 \\
-1.231 &  172.118 &  4978.11 & 831.86  &  2803.63  &  4.015e+22 &  4.292e+18 &  9.292e-05 \\
-0.885 &  118.163 &  5236.73 & 484.64  &  3021.62  &  6.006e+22 &  7.249e+18 &  1.377e-04 \\
-0.585 &   70.483 &  5566.51 & 172.68  &  3215.84  &  8.168e+22 &  1.287e+19 &  1.872e-04 \\
-0.273 &   25.737 &  6083.30 & -155.86 &  3402.02  &  1.031e+23 &  3.319e+19 &  2.363e-04 \\
 0.111 &  -13.797 &  6888.11 & -557.48 &  3572.90  &  1.177e+23 &  1.461e+20 &  2.700e-04 \\
 0.512 &  -41.669 &  7849.90 & -959.63 &  3649.49  &  1.198e+23 &  6.309e+20 &  2.761e-04 \\
 0.909 &  -65.000 &  8721.04 & -1326.95 & 3702.02  &  1.190e+23 &  1.835e+21 &  2.771e-04 \\

\hline
\end{tabular}%
}
\end{table}

\section{Uncertainties in the inverted atmospheric parameters.}

In this appendix we provide an estimate of the uncertainties in tabulated form (indicated in Table~\ref{error:cool}, Table~\ref{error:hot}, Table~\ref{error:flash} and Table~\ref{error:qs}), for the main atmospheric parameters and for each of the proposed atmospheric models, by using the inversion uncertainties as defined in equation 42 of \citet{Iniesta_2016}. These are computed per height on the basis of the response functions and do not take into account that a change in a node will affect the atmosphere at all heights between such node and the next. Further, these uncertainties best apply to the inversion but do not reflect the selection procedure performed to select only models that also lead to \ion{Mg}{ii}~h\&k profiles that reproduce the observations.

\begin{table}[]
\centering
\caption{Inversion uncertainties for the cool penumbra.}
\label{error:cool}
\begin{tabular}{cccc}
\hline
$\log\tau_{5000}$ & Height (km) & $\sigma_{\mathrm{T}}$ (K) & $\sigma_{\mathrm{V}_\mathrm{LOS}}$ (m s$^{-1})$ \\
\midrule
-8.000 & 1725.29 & 2.41e+06 & Unconstrained \\
-7.767 & 1724.87 & 75687 & Unconstrained \\
-7.744 & 1724.28 & 75519 & Unconstrained \\
-7.716 & 1723.68 & 1.16e+06 & Unconstrained \\
-7.688 & 1723.20 & 3.89e+06 & Unconstrained \\
-7.670 & 1722.95 & 75520 & 3.87582e+09 \\
-7.650 & 1722.67 & 145298 & 2.74062e+09 \\
-7.619 & 1722.27 & 65477 & 4.27003e+06 \\
-7.569 & 1721.59 & 49919 & 604313. \\
-7.493 & 1720.46 & 74840 & 235983. \\
-7.367 & 1718.18 & 67594 & 95582.8 \\
-7.181 & 1713.48 & 52340 & 29101.7 \\
-6.970 & 1705.65 & 33572 & 26215.9 \\
-6.749 & 1693.28 & 20767 & 15094.3 \\
-6.526 & 1675.11 & 8457 & 8927.92 \\
-6.309 & 1647.53 & 1357 & 4454.89 \\
-6.117 & 1591.18 & 263 & 1428.14 \\
-5.941 & 1483.93 & 19 & 467.590 \\
-5.813 & 1387.89 & 1.0 & 392.19 \\
-5.718 & 1314.75 & 1.1 & 287.53 \\
-5.629 & 1247.10 & 1.4 & 246.31 \\
-5.545 & 1184.27 & 112 & 243.99 \\
-5.462 & 1125.30 & 1.0 & 277.88 \\
-5.378 & 1068.84 & 1.1 & 387.79 \\
-5.292 & 1014.79 & 1.0 & 688.50 \\
-5.204 & 963.200 & 41 & 1587.36 \\
-5.108 & 913.011 & 1.0 & 936.64 \\
-5.005 & 864.302 & 19 & 613.87 \\
-4.898 & 817.129 & 1.0 & 579.29 \\
-4.788 & 771.021 & 20 & 625.75 \\
-4.674 & 725.797 & 1.0 & 788.91 \\
-4.561 & 683.237 & 106 & 911.00 \\
-4.447 & 647.910 & 3.5 & 1182.19 \\
-4.328 & 619.563 & 217 & 2882.66 \\
-4.195 & 593.137 & 8.0 & 3082.49 \\
-4.034 & 565.973 & 60 & 4851.83 \\
-3.831 & 534.992 & 120 & 4785.97 \\
-3.561 & 496.785 & 73 & 2880.21 \\
-3.204 & 448.469 & 60 & 1435.90 \\
-2.814 & 396.318 & 35 & 1023.88 \\
-2.420 & 343.362 & 21 & 1023.33 \\
-2.024 & 289.566 & 15 & 1202.77 \\
-1.626 & 235.260 & 13 & 1533.09 \\
-1.231 & 180.041 & 14 & 1894.30 \\
-0.8850 & 129.745 & 22 & 3328.24 \\
-0.5850 & 84.0550 & 30 & 6020.62 \\
-0.2730 & 34.5280 & 129 & 38906.0 \\
0.1110 & -19.8620 & 7561 & 5.93534e+07 \\
0.5120 & -59.3100 & 26951 & 2.22578e+09 \\
0.9090 & -91.2460 & 987007 & Unconstrained \\
\hline
\end{tabular}%
\end{table}

\begin{table}[]
\centering
\caption{Inversion uncertainties for the hot penumbra.}
\label{error:hot}
\begin{tabular}{cccc}
\hline
$\log\tau_{5000}$ & Height (km) & $\sigma_{\mathrm{T}}$ (K) & $\sigma_{\mathrm{V}_\mathrm{LOS}}$ (m s$^{-1})$ \\
\midrule
-8.000 & 1744.15 & 118612 & Unconstrained \\
-7.767 & 1743.71 & 118623 & Unconstrained \\
-7.744 & 1743.10 & 118562 & Unconstrained \\
-7.716 & 1742.47 & 5.95e+06 & Unconstrained \\
-7.688 & 1741.96 & 6.07e+06 & Unconstrained \\
-7.670 & 1741.68 & 37496 & Unconstrained \\
-7.650 & 1741.38 & 8490 & Unconstrained \\
-7.619 & 1740.93 & 5441 & 6.47402e+10 \\
-7.569 & 1740.18 & 7751 & 1.47092e+06 \\
-7.493 & 1738.93 & 8143 & 318707. \\
-7.367 & 1736.37 & 117993 & 123977. \\
-7.181 & 1731.04 & 116613 & 45272.1 \\
-6.970 & 1721.97 & 116031 & 28301.1 \\
-6.749 & 1707.46 & 25796 & 16165.0 \\
-6.526 & 1686.14 & 11798 & 9065.94 \\
-6.309 & 1658.22 & 9115 & 5749.42 \\
-6.117 & 1624.85 & 1358 & 3411.52 \\
-5.941 & 1574.10 & 338 & 1525.96 \\
-5.813 & 1513.17 & 191 & 1165.83 \\
-5.718 & 1450.40 & 245 & 652.720 \\
-5.629 & 1377.41 & 181 & 381.230 \\
-5.545 & 1297.20 & 36 & 247.800 \\
-5.462 & 1213.89 & 69 & 195.230 \\
-5.378 & 1131.33 & 31 & 209.930 \\
-5.292 & 1053.38 & 36 & 339.220 \\
-5.204 & 981.679 & 39 & 848.060 \\
-5.108 & 915.365 & 37 & 779.030 \\
-5.005 & 853.993 & 71 & 626.160 \\
-4.898 & 797.356 & 65 & 692.850 \\
-4.788 & 744.541 & 73 & 777.870 \\
-4.674 & 695.532 & 154 & 1008.00 \\
-4.561 & 653.802 & 1.0 & 1224.20 \\
-4.447 & 622.481 & 437 & 2752.99 \\
-4.328 & 597.786 & 362 & 5826.92 \\
-4.195 & 574.639 & 422 & 9630.23 \\
-4.034 & 549.903 & 354 & 5907.76 \\
-3.831 & 520.729 & 235 & 2768.92 \\
-3.561 & 483.854 & 88 & 1296.33 \\
-3.204 & 436.227 & 56 & 771.810 \\
-2.814 & 384.834 & 31 & 688.420 \\
-2.420 & 333.047 & 17 & 684.780 \\
-2.024 & 280.491 & 11 & 792.890 \\
-1.626 & 226.800 & 9.7 & 1011.89 \\
-1.231 & 172.810 & 11 & 1160.48 \\
-0.8850 & 123.237 & 17 & 1871.28 \\
-0.5850 & 77.3340 & 23 & 3291.10 \\
-0.2730 & 28.9240 & 94 & 20742.6 \\
0.1110 & -14.2830 & 2464 & 4.19414e+06 \\
0.5120 & -39.5560 & 7167 & 2.97225e+11 \\
0.9090 & -57.3250 & 83506 & Unconstrained \\
\hline
\end{tabular}%
\end{table}

\begin{table}[]
\centering
\caption{Inversion uncertainties for the umbral flash.}
\label{error:flash}
\begin{tabular}{cccc}
\hline
$\log\tau_{5000}$ & Height (km) & $\sigma_{\mathrm{T}}$ (K) & $\sigma_{\mathrm{V}_\mathrm{LOS}}$ (m s$^{-1})$ \\
\midrule
-8.000 & 1654.25 & 1.12e+07 & Unconstrained \\
-7.767 & 1653.79 & 1.33e+07 & Unconstrained \\
-7.744 & 1653.15 & 8.61e+06 & Unconstrained \\
-7.716 & 1652.47 & 2787 & Unconstrained \\
-7.688 & 1651.90 & 2787 & Unconstrained \\
-7.670 & 1651.58 & 2787 & 2.44988e+10 \\
-7.650 & 1651.24 & 2787 & 2.44988e+10 \\
-7.619 & 1650.72 & 3057 & 3.03871e+09 \\
-7.569 & 1649.83 & 459809 & 2.86737e+09 \\
-7.493 & 1648.34 & 1.77e+06 & Unconstrained \\
-7.367 & 1645.26 & 3006 & 149073. \\
-7.181 & 1638.77 & 3362 & 169516. \\
-6.970 & 1627.63 & 3772 & 141635. \\
-6.749 & 1609.70 & 63243 & 127651. \\
-6.526 & 1583.25 & 50463 & 49830.9 \\
-6.309 & 1548.19 & 96373 & 24696.8 \\
-6.117 & 1510.21 & 33227 & 15868.9 \\
-5.941 & 1469.70 & 3963 & 10947.3 \\
-5.813 & 1431. & 3081 & 10031.6 \\
-5.718 & 1387.96 & 814 & 5475.48 \\
-5.629 & 1325.96 & 830 & 2644.51 \\
-5.545 & 1239.27 & 430 & 1240.73 \\
-5.462 & 1131.06 & 29 & 682.570 \\
-5.378 & 1015.39 & 108 & 581.610 \\
-5.292 & 908.743 & 66 & 847.070 \\
-5.204 & 817.634 & 61 & 1299.57 \\
-5.108 & 740.030 & 62 & 1086.93 \\
-5.005 & 673.424 & 121 & 929.480 \\
-4.898 & 618.331 & 166 & 1076.94 \\
-4.788 & 577.021 & 261 & 1292.72 \\
-4.674 & 548.312 & 297 & 1894.41 \\
-4.561 & 527.361 & 536 & 2576.64 \\
-4.447 & 508.172 & 780 & 2569.30 \\
-4.328 & 487.696 & 352 & 2104.04 \\
-4.195 & 464.538 & 594 & 1806.49 \\
-4.034 & 438.912 & 586 & 2227.58 \\
-3.831 & 410.283 & 548 & 3065.70 \\
-3.561 & 376.920 & 875 & 4482.95 \\
-3.204 & 337.707 & 109 & 1468.87 \\
-2.814 & 297.305 & 38 & 3782.62 \\
-2.420 & 254.185 & 30 & 3581.27 \\
-2.024 & 208.486 & 50 & 12757.9 \\
-1.626 & 161.803 & 36 & 7170.65 \\
-1.231 & 115.809 & 37 & 5882.61 \\
-0.8850 & 77.2130 & 52 & 8824.87 \\
-0.5850 & 46.5870 & 73 & 15058.8 \\
-0.2730 & 19.1100 & 301 & 82536.7 \\
0.1110 & -14.6000 & 1834 & 2.29667e+08 \\
0.5120 & -60.8800 & 5302 & 1.90139e+11 \\
0.9090 & -104.583 & 132329 & Unconstrained \\

\hline
\end{tabular}
\end{table}

\begin{table}[]
\centering
\caption{Inversion uncertainties for the quiet surrounding atmosphere.}
\label{error:qs}
\begin{tabular}{cccc}
\hline
$\log\tau_{5000}$ & Height (km) & $\sigma_{\mathrm{T}}$ (K) & $\sigma_{\mathrm{V}_\mathrm{LOS}}$ (m s$^{-1})$ \\
\midrule
-7.767 & 1865.24 & 143691. & Unconstrained \\
-7.716 & 1863.97 & 143691. & Unconstrained \\
-7.650 & 1862.84 & 143691. & Unconstrained \\
-7.619 & 1862.37 & 143691. & Unconstrained \\
-7.569 & 1861.59 & 143691. & 8.29600e+06 \\
-7.367 & 1857.62 & 132.1 & 190944. \\
-7.181 & 1852.09 & 132.1 & 83749.6 \\
-6.970 & 1842.64 & 132.2 & 44219.1 \\
-6.749 & 1827.33 & 132.3 & 23202.3 \\
-6.526 & 1804.38 & 18867.6 & 12134.4 \\
-6.309 & 1773.80 & 11431.4 & 7238.16 \\
-6.117 & 1739.14 & 2460.6 & 4711.14 \\
-5.941 & 1699.47 & 1874.51 & 3069.99 \\
-5.813 & 1663.14 & 64.5 & 3308.33 \\
-5.718 & 1630.67 & 62.7 & 2420.50 \\
-5.629 & 1594.55 & 59.7 & 1701.09 \\
-5.545 & 1553.20 & 282.1 & 1147.10 \\
-5.462 & 1505.07 & 108.2 & 743.14 \\
-5.378 & 1448.03 & 140. & 471.51 \\
-5.292 & 1380.79 & 113.6 & 311.25 \\
-5.204 & 1301.96 & 77.9 & 225.07 \\
-5.108 & 1210.53 & 30.7 & 210.80 \\
-5.005 & 1109.75 & 18.3 & 331.82 \\
-4.898 & 1007.87 & 37.1 & 828.04 \\
-4.788 & 913.033 & 29.5 & 360.19 \\
-4.674 & 827.766 & 48.2 & 340.51 \\
-4.561 & 751.574 & 81.7 & 454.13 \\
-4.447 & 689.063 & 56.9 & 441.29 \\
-4.328 & 642.639 & 53.6 & 1324.25 \\
-4.195 & 606.191 & 76.8 & 2909.85 \\
-4.034 & 573.350 & 58.7 & 5062.82 \\
-3.831 & 539.644 & 59.2 & 2728.15 \\
-3.561 & 500.174 & 123.6 & 1123.31 \\
-3.204 & 450.950 & 76.3 & 667.40 \\
-2.814 & 397.912 & 42.7 & 620.17 \\
-2.420 & 343.609 & 23 & 641.39 \\
-2.024 & 288.647 & 13.7 & 736.75 \\
-1.626 & 231.283 & 11.4 & 955.91 \\
-1.231 & 172.118 & 12.6 & 1368.50 \\
-0.8850 & 118.163 & 18.3 & 2424.72 \\
-0.5850 & 70.4830 & 27.2 & 4822.81 \\
-0.2730 & 25.7370 & 126.2 & 29102.8 \\
0.1110 & -13.7970 & 3107.6 & 1.8e+06 \\
0.5120 & -41.6690 & 12459.6 & Unconstrained \\
0.9090 & -65.0000 & 64260.6 & Unconstrained \\

\hline
\end{tabular}%
\end{table}

\end{appendix}
\end{document}